\documentclass{vgtc}                          

\ifpdf
  \pdfoutput=1\relax                   
  \usepackage{graphicx}                
  \DeclareGraphicsExtensions{.pdf,.png,.jpg,.jpeg} 
\else
  \usepackage{graphicx}                
  \DeclareGraphicsExtensions{.eps}     
\fi%

\graphicspath{{figures/}{pictures/}{images/}{./}} 

\usepackage{microtype}                 
\PassOptionsToPackage{warn}{textcomp}  
\usepackage{textcomp}                  
\usepackage{mathptmx}                  
\usepackage{times}                     
\usepackage{cite}                      
\usepackage{tabu}                      
\usepackage{booktabs}                  
\usepackage{bm}
\usepackage{amssymb}
\usepackage{multirow}

\usepackage{comment}
\usepackage{amsmath}
\DeclareMathOperator*{\cd}{CD}

\onlineid{0}

\vgtccategory{IEEE VIS Short Paper}

\vgtcinsertpkg

\title{A Comparative Study of Neural Surface Reconstruction for \\ Scientific Visualization}

\author{Siyuan Yao\thanks{e-mail: syao2@nd.edu}\\ %
        \scriptsize University of Notre Dame %
\and Weixi Song\thanks{e-mail: song.wx@whu.edu.cn}\\ %
     \scriptsize Wuhan University %
\and Chaoli Wang\thanks{e-mail: chaoli.wang@nd.edu}\\ %
     \scriptsize University of Notre Dame %
}

\abstract{
This comparative study evaluates various neural surface reconstruction methods, particularly focusing on their implications for scientific visualization through reconstructing 3D surfaces via multi-view rendering images. We categorize ten methods into neural radiance fields and neural implicit surfaces, uncovering the benefits of leveraging distance functions (i.e., SDFs and UDFs) to enhance the accuracy and smoothness of the reconstructed surfaces. Our findings highlight the efficiency and quality of NeuS2 for reconstructing closed surfaces and identify NeUDF as a promising candidate for reconstructing open surfaces despite some limitations. By sharing our benchmark dataset, we invite researchers to test the performance of their methods, contributing to the advancement of surface reconstruction solutions for scientific visualization.
} 

\begin{document}


\vspace{-0.05in}
\firstsection{Introduction}

\maketitle



Multi-view 3D surface reconstruction represents a pivotal progress in computer vision, offering a methodology for synthesizing 3D geometric models from multiple 2D images captured from varying camera views. 
This research involves sophisticated algorithms integrating these disparate views to form a cohesive 3D representation of the observed object or scene. 

In scientific visualization, this research direction is advantageous when only surface rendering images, rather than the original surface or volume data, are available. 
Such a scenario is common, as original volume data and extracted surfaces are not always publicly disclosed. 
With a small set of rendering images of the surface captured from different viewpoints, we can faithfully reconstruct the 3D surface, enabling examination from arbitrary angles. 
Additionally, with the reconstructed surface, we can adjust lighting and rendering parameters to enhance surface details or present the surface in a preferred manner. 
For large volumetric datasets, surface rendering images can be remotely produced by high-performance computing (HPC) clusters and efficiently transferred to local users. The corresponding surfaces can be locally reconstructed in minutes using the resulting rendering images instead of the volume data. 

Yet, a significant gap remains in our understanding of how these reconstruction methods perform when applied to scientific datasets. 
In scientific visualization, various surface types present unique challenges for 3D surface reconstruction, and no single method is guaranteed to work for every dataset.
Special care must be taken to account for data characteristics and make necessary tradeoffs to meet the demands of professional users or the general public.
  
For example, isosurfaces can be complex and exhibit occlusions, with inner surfaces visible only from certain viewpoints. In such a case, a method that excels in occlusion tolerance or maintains good surface consistency may be preferable over efficiency. 
Conversely, for remote reconstruction from HPC clusters, users might prioritize more efficient methods to quickly obtain the surface.

Different surface characteristics also play a crucial role. 
Although isosurfaces usually include open and closed surfaces, applying closed surface reconstruction methods can still produce reasonable results. 
However, this approach is not effective for stream surfaces. 
Using closed surface reconstruction methods on stream surfaces, which are often open, can lead to extra faces or incorrect face connections, making the reconstructed surface very different from the rendered appearance. 
Even with the existing open surface methods, we must balance accuracy, smoothness, and speed.
  
To bridge this gap, we introduce a benchmark dataset to thoroughly compare ten state-of-the-art surface reconstruction techniques. 
Our dataset encompasses nine isosurfaces and stream surfaces, ranging from simple to complex, including closed and open surfaces, each with distinct characteristics. 
Through this comprehensive study, we shed light on the performance of these methods, offering valuable insights and recommendations. 

\vspace{-0.05in}
\section{Related Work}

{\bf Data reconstruction via deep learning.}
Deep learning has accomplished many scientific visualization tasks, including data resconstruction~\cite{Wang-DL4SciVis}, which focuses on restoring data from visual or incomplete sources. 
Examples include~\cite{Han-CGA19,Gu-VFR-UFD,Han-VCNet,Weiss-TVCG21,Weiss-TVCG22}. 
%
%
Nevertheless, in scientific visualization, no reconstruction technique has been developed that learns surface representations from a collection of 2D rendered images taken from various angles, as opposed to utilizing volumetric data~\cite{Han-FlowNet} or surface meshes~\cite{Han-SurfNet}.

{\bf Explicit surface reconstruction.}
Reconstructing 3D surfaces from multi-view 2D images is common in computer vision and computer graphics. 
Typically, 3D surfaces can be represented explicitly as {\em voxels}~\cite{Choy-ECCV16, Xie-ICCV19, Li-TVCG24}, {\em point clouds}~\cite{Fan-CVPR17, Mandikal-BMVC18, Lin-AAAI18, Achlioptas-PMLR18}, or {\em meshes}~\cite{Groueix-CVPR18, Wang-ECCV18, Wen-ICCV19}, and surface reconstruction is often performed using techniques like marching cubes~\cite{Lorensen-SIGGRAPH87}, ball pivoting~\cite{Bernardini-TVCG99}, and Poisson reconstruction~\cite{Kazhdan-SGP06}. 
Due to the discrete nature of these representations, surface reconstruction from these fully explicit representations usually yields noisy and incomplete results.

{\bf Implicit surface reconstruction.}
The recent surge of neural rendering frameworks~\cite{Mildenhall-ECCV20, Yariv-NeurIPS20, Yariv-NeurIPS21} and implicit surface representations~\cite{Mescheder-CVPR19, Park-CVPR19, Remelli-NeurIPS20, Guillard-ECCV22} has propelled advances in multi-view 3D surface reconstruction with considerable improvements in accuracy, consistency, and flexibility. 
Although neural rendering frameworks like NeRF~\cite{Mildenhall-ECCV20} can extract surfaces from learned density fields, the results tend to be rough and irregular due to the lack of geometric constraints. 
Multi-view surface reconstruction methods, such as VolSDF~\cite{Yariv-NeurIPS21} and NeuS~\cite{Wang-NeurIPS21}, employ the {\em signed distance function} (SDF) for implicit surface representation. 
NeuS2~\cite{Wang-ICCV23} and Neuralangelo~\cite{Li-CVPR23} significantly improve the training speed of NeuS by integrating {\em multiresolution hash encoding} (MHE). 
NeAT~\cite{Meng-CVPR23} introduces a validation network for open surface reconstruction. 
In contrast, NeUDF~\cite{Liu-CVPR23} and NeuralUDF~\cite{Long-CVPR23} employ an {\em unsigned distance field} (UDF) to represent open surfaces and achieve high-quality reconstruction. 
We assess and benchmark various neural surface reconstruction methods, discussing their adaptation to isosurfaces and stream surfaces commonly found in scientific visualization, offering recommendations, and outlining remaining challenges. 

\vspace{-0.05in}
\section{Radiance Fields and Implicit Surfaces}

{\bf Neural radiance fields.}
Neural radiance field (NeRF)~\cite{Mildenhall-ECCV20}, a seminal work on view synthesis, uses a deep, fully connected network that ingests a continuous 5D coordinate (including 3D spatial location and 2D viewing angle) and outputs the corresponding volume density and view-dependent emitted radiance. 
This method synthesizes views by iterating 5D coordinates along camera rays and employs the standard volume rendering process to produce the novel view. 
One of its most striking features is NeRF's simplicity, i.e., using a {\em multilayer perceptron} (MLP) to process these coordinates and output density and color. 
NeRF can represent detailed scene geometry with complex occlusions and even convert the radiance field to a mesh using marching cubes. 
However, the vanilla NeRF has limitations such as slow training and rendering speeds, inability to represent dynamic scenes, and ``baking in" lighting. 

TensoRF~\cite{Chen-ECCV22} pushes the boundaries of 3D scene reconstruction regarding efficiency and fidelity. 
In contrast to NeRF (which depends entirely on the use of MLPs for scene rendering) and Plenoxels~\cite{Fridovich-Keil-CVPR22} (which relies on a fully explicit sparse voxel grid for volume representation), TensoRF conceptualizes the entire volume of a scene as a 4D tensor, including the feature channel as one dimension. 
Specifically, it involves decomposing the scene's tensor into several compact, low-rank tensor components, enabling a more refined and efficient model representation using a significantly smaller MLP than NeRF. 
Compared with Plenoxels, TensoRF offers a substantial reduction in memory usage due to its efficient tensor factorization for handling the feature grid, which enhances scene modeling and allows for more efficient and accurate rendering.

Using MHE, Instant-NGP~\cite{Mueller-SIGGRAPH22} replaces many parameters in the NeRF network with a smaller one, supplemented by a set of trainable encoding parameters. 
These encoding parameters are stored at the vertices of multiple grid layers, facilitating the learning of scene details at various resolutions. 
A key feature of Instant-NGP is its ability to achieve almost-instant training of neural graphics primitives using a single GPU. 
Moreover, Instant-NGP's architecture allows real-time training progress on various datasets, supporting complex scenes. 
Its implementation in a single CUDA kernel, referred to as ``fully-fused MLP," results in a tenfold efficiency improvement compared to the original NeRF implementation. 

{\bf Neural implicit surfaces.}
Differentiable volumetric renderer (DVR)~\cite{Niemeyer-CVPR20} and implicit differentiable renderer (IDR)~\cite{Yariv-NeurIPS20} significantly advance neural implicit surface reconstruction from multi-view 3D images, laying the groundwork for subsequent development. 
NeuS~\cite{Wang-NeurIPS21} further eliminates the need for mask supervision in DVR and IDR. 
Using an SDF and a new volume rendering method for training, NeuS allows for more accurate surface reconstruction, especially in complex structures and self-occlusion cases. 
NeuS2~\cite{Wang-ICCV23} and Neuralangelo~\cite{Li-CVPR23}, both implementations of NeuS within the Instant-NGP framework, permit instantaneous training. 
However, this comes at the cost of noise artifacts in both approaches.

NeUDF~\cite{Liu-CVPR23} and NeuralUDF~\cite{Long-CVPR23} focus on reconstructing surfaces with open boundaries and arbitrary topologies, utilizing UDFs.  
In NeUDF, a differentiable volume rendering framework is introduced to predict UDFs from input images, emphasizing the need for a specialized rendering procedure and a point sampling strategy tailored for UDFs. 
NeuralUDF overcomes the challenges of locating the zero level set for a UDF field, which is crucial for accurately defining open surfaces. 
Leveraging a modified visibility indicator function and optimizing the UDF fields, NeuralUDF can accurately recover objects with open surfaces from 2D images.

\vspace{-0.05in}
\section{Comparison and Discussion}

For a more nuanced comparative analysis, we categorize ten methods into three distinct groups: 
NeRF (including NeRF~\cite{Mildenhall-ECCV20}, TensoRF~\cite{Chen-ECCV22}, and Instant-NGP~\cite{Mueller-SIGGRAPH22}), 
neural implicit closed surface (featuring IDR~\cite{Yariv-NeurIPS20}, NeuS~\cite{Wang-NeurIPS21}, NeuS2~\cite{Wang-ICCV23}, and Neuralangelo~\cite{Li-CVPR23}), and 
neural implicit open surface (comprising NeAT~\cite{Meng-CVPR23}, NeUDF~\cite{Liu-CVPR23}, and NeuralUDF~\cite{Long-CVPR23}). 
We include NeuS in all comparisons as it stands out as the most representative work. 
NeuS is closely related to the methods in the three groups and effectively highlights their distinctions. 
This section compares neural renderings and reconstructed surfaces in static images. 
The accompanying video compares the results across the full 360-degree view. 

\vspace{-0.05in}
\subsection{Datasets, Training, and Metrics}

We used nine datasets in our experiments, with each group comparison using three datasets to showcase the differences. 
Table~\ref{tab:dataset} lists the volumetric datasets, spatial resolutions, and surface information. 
The five critical points and solar plume datasets use stream surfaces, while all others use isosurfaces.
For the isosurface, we selected an isovalue leading to a salient isosurface covering the domain well. 
Note that for the supernova dataset, the isosurface exhibits a complex, multi-layered structure, including inner parts hidden from any viewing angle. 
We use this challenging case to evaluate the robustness of surface reconstruction methods and the consistency of their reconstructed surfaces. 
For the stream surface, we picked a seeding curve to yield a prominent stream surface. 
For each dataset, we used 42 rendering images from evenly-placed sample views for training and synthesized 181 new
views during inference.
The image resolution is 1024$\times$1024 for both training and inference. 
We also reconstructed the underlying surface through the learned density field, SDF, or UDF. 
All methods were trained on a single NVIDIA V100 GPU using the default configurations suggested in the respective works. 
Then, we rendered the reconstructed isosurface or stream surface as usual using the same 181 views. 
The details of training parameters are presented in Table~\ref{tab:training}.

\begin{table}[htb]
  \vspace{-0.05in}
  \caption{Datasets, spatial resolutions, and surface information.}
  \vspace{-0.05in}
  \centering
   \resizebox{\columnwidth}{!}{
    \begin{tabular}{c|c|ccc}
     &              & dimension                & surface & \# original  \\ 
     & dataset & $(x \times y \times z)$ & type     & vertices     \\ \hline
     neural & Nyx & 256$\times$256$\times$256 & isosurface & 211,910  \\
     radiance & Tangaroa & 300$\times$180$\times$120 & isosurface & 162,187  \\
     field & vortex & 128$\times$128$\times$128 & isosurface & 153,574  \\ \hline
     neural & aorta & 512$\times$666$\times$251 & isosurface & 949,999  \\
     closed & combustion & 480$\times$720$\times$120 & isosurface & 1,945,531  \\
     surface & supernova & 432$\times$432$\times$432 & isosurface & 2,396,278  \\ \hline
     neural & five critical points & 51$\times$51$\times$51 & stream surface & 2,072  \\
     open & Marschner-Lobb & 256$\times$256$\times$256 & isosurface & 497,108  \\
     surface & solar plume & 126$\times$126$\times$512 & stream surface & 85,863  \\ 
     \end{tabular}
    }
    \label{tab:dataset}
 \vspace{-0.1in}    
\end{table}

\begin{table}[htb]
   \vspace{-0.05in}
    \caption{Training details of the three groups of methods.}
    \vspace{-0.05in}
    \centering
    \scalebox{0.75}{
    \begin{tabular}{c|cccc}
    & \multicolumn{4} {c} {neural radiance field} \\ \hline
     method & NeRF & TensoRF & Instant-NGP & NeuS \\ \hline
     framework & PyTorch & PyTorch & CUDA & PyTorch \\
     learning rate & $5\times10^{-4}$ & $1\times10^{-3}$ & $1\times10^{-3}$ & $5\times10^{-4}$ \\
     \# iterations & 100k & 30k & 35k & 300k \\
    \end{tabular}
    }
    \scalebox{0.75}{
    \begin{tabular}{c|cccc}
    & \multicolumn{4} {c} {neural closed surface} \\ \hline
     method & IDR & NeuS & NeuS2 & Neuralangelo \\ \hline
     framework & PyTorch & PyTorch & CUDA & PyTorch \\
     learning rate & $1\times10^{-4}$ & $5\times10^{-4}$ & $1\times10^{-2}$ & $1\times10^{-3}$ \\
     \# iterations & 1000k & 300k & 15k & 500k \\
    \end{tabular}
    }
    \scalebox{0.75}{
    \begin{tabular}{c|cccc}
    & \multicolumn{4} {c} {neural open surface} \\ \hline
     method & NeuS & NeAT & NeUDF & NeuralUDF \\ \hline
    framework & PyTorch & PyTorch & PyTorch & PyTorch \\
    learning rate & $5\times10^{-4}$ & $1\times10^{-4}$ & $2\times10^{-4}$ & $5\times10^{-4}$ \\
    \# iterations & 300k & 1000k & 400k & 300k \\
    \end{tabular}
    }
    \label{tab:training}
 \vspace{-0.1in}
\end{table}

For quantitative comparison, we used peak signal-to-noise ratio (PSNR) and learned perceptual image patch similarity (LPIPS)~\cite{Zhang-CVPR18} to evaluate the quality of inferred neural rendering images. 
We utilized the Chamfer distance (CD)~\cite{Barrow-IJCAI77} to compute the difference between reconstructed surface $\hat{S}$ and ground truth (GT) surface $S$
\begin{equation}
  \begin{array}{l}
    \displaystyle d_{\cd}(S,\hat{S})=\sum_{v\in S} \min_{\hat{v}\in \hat{S}}  ||v-\hat{v}||^2_2 + \sum_{\hat{v}\in \hat{S}} \min_{v\in S}  ||\hat{v}-v||^2_2.
    \label{eqn:nerf-color-integral}
  \end{array}
\end{equation}
We reported CD in the above two terms rather than their summation. 
The first term measures the distance from GT surface vertices $v$ to the nearest vertices on the reconstructed surface $\hat{S}$, and the second term from reconstructed surface vertices $\hat{v}$ to the nearest vertices on the GT surface $S$. 
This separation provides a clearer assessment of reconstruction accuracy.

\begin{figure}[htb]
 \vspace{-.1in} 
  \begin{center}
  $\begin{array}{c@{\hspace{0.02in}}c@{\hspace{0.02in}}c@{\hspace{0.02in}}c@{\hspace{0.02in}}c}
  \includegraphics[width=0.1875\linewidth]{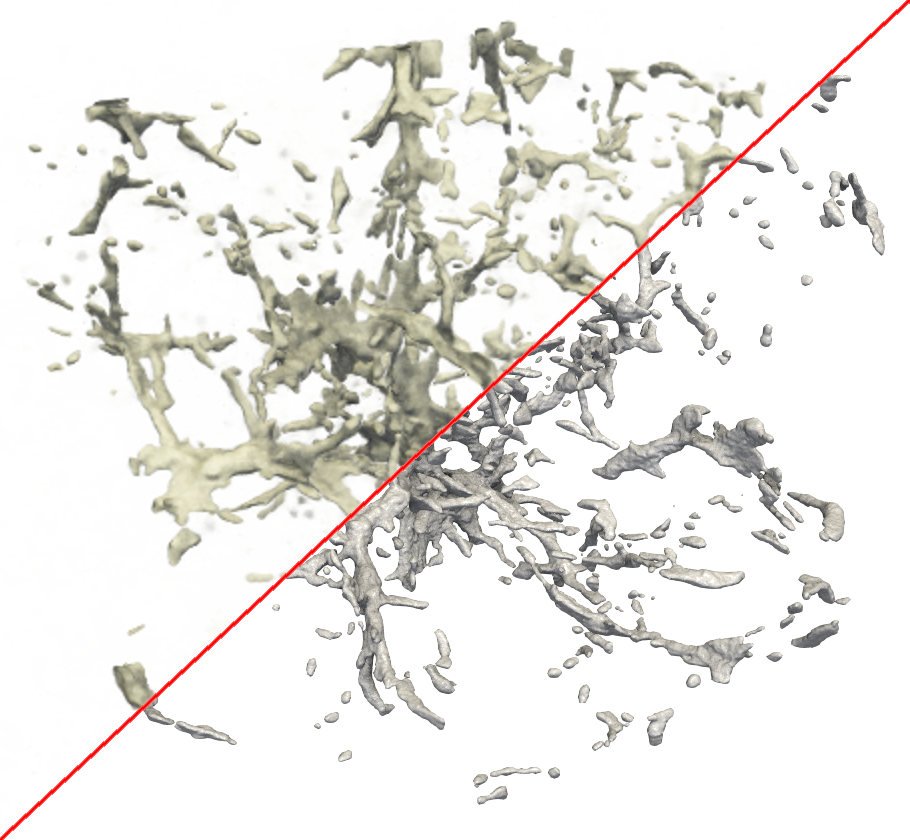}&
  \includegraphics[width=0.1875\linewidth]{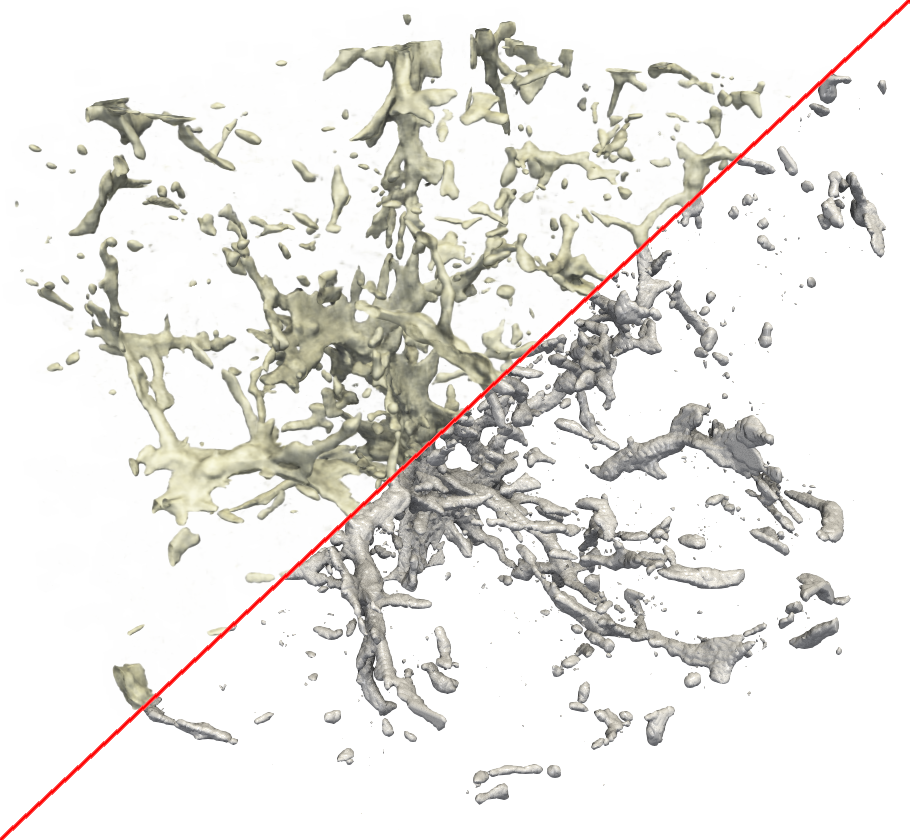}&
  \includegraphics[width=0.1875\linewidth]{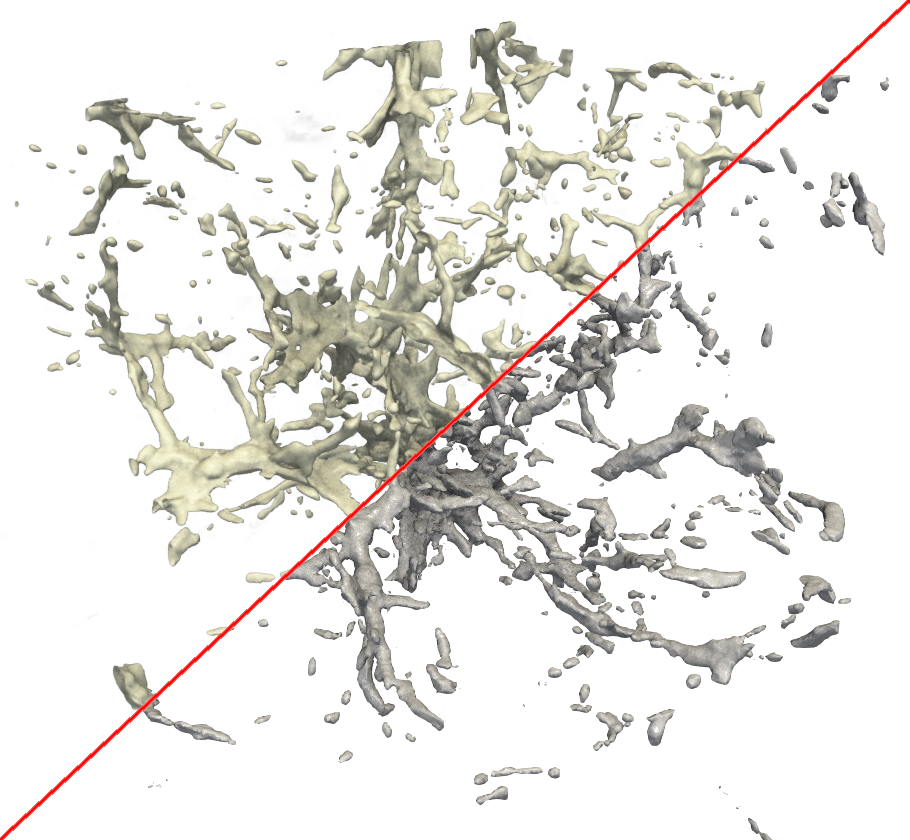}&
  \includegraphics[width=0.1875\linewidth]{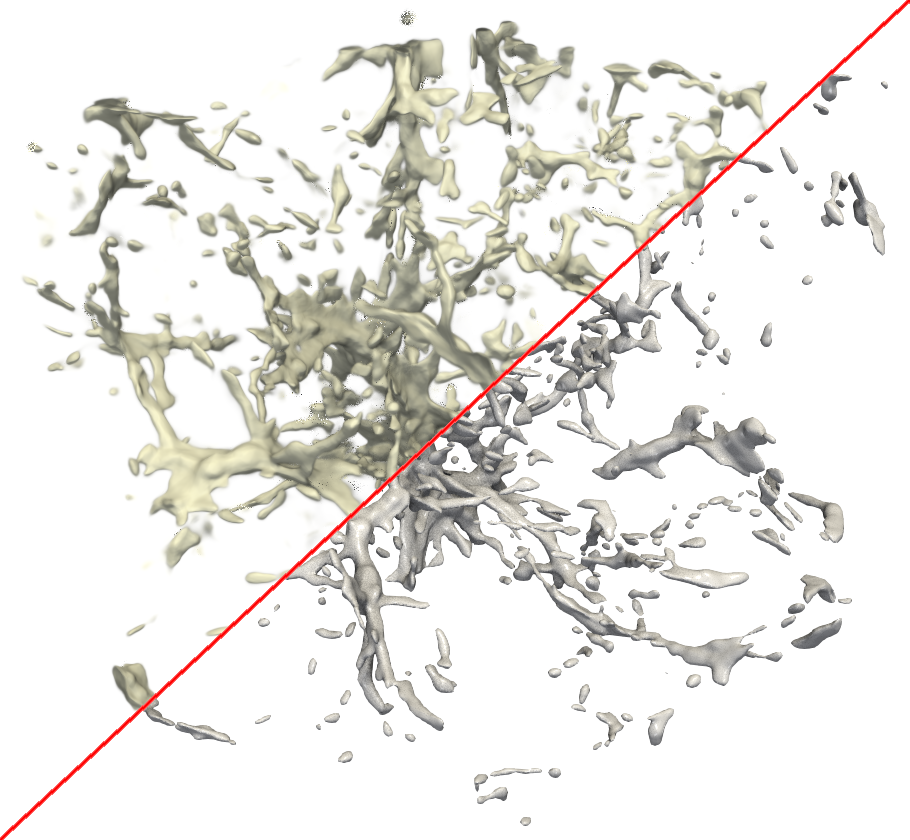}&
  \includegraphics[width=0.1875\linewidth]{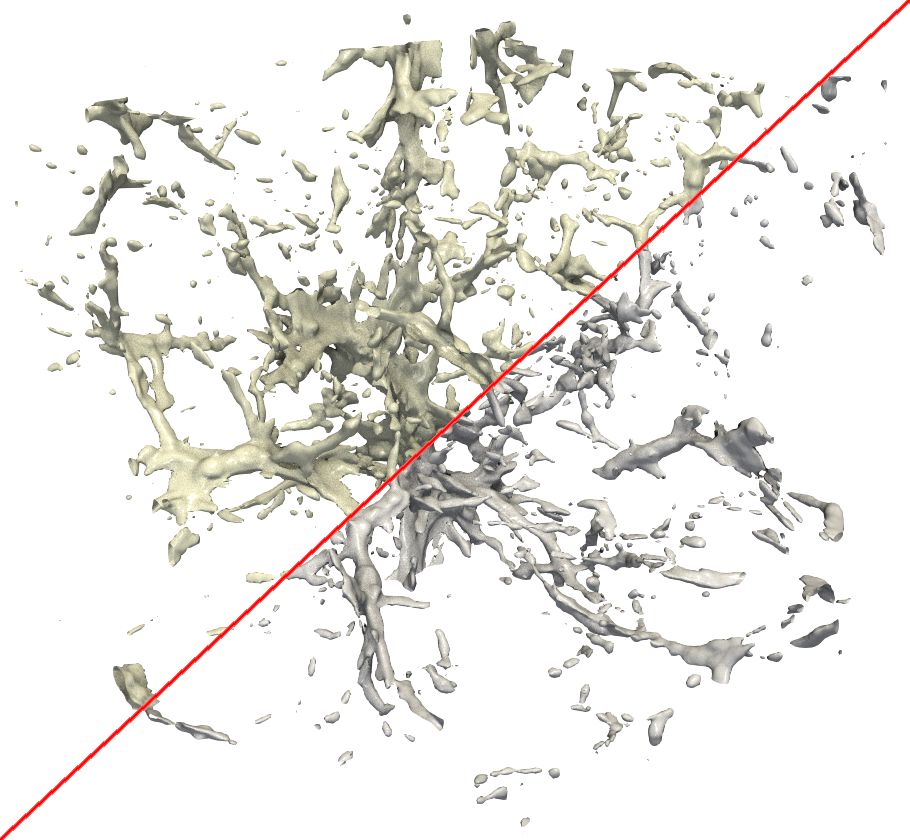}\\
  \includegraphics[width=0.1875\linewidth]{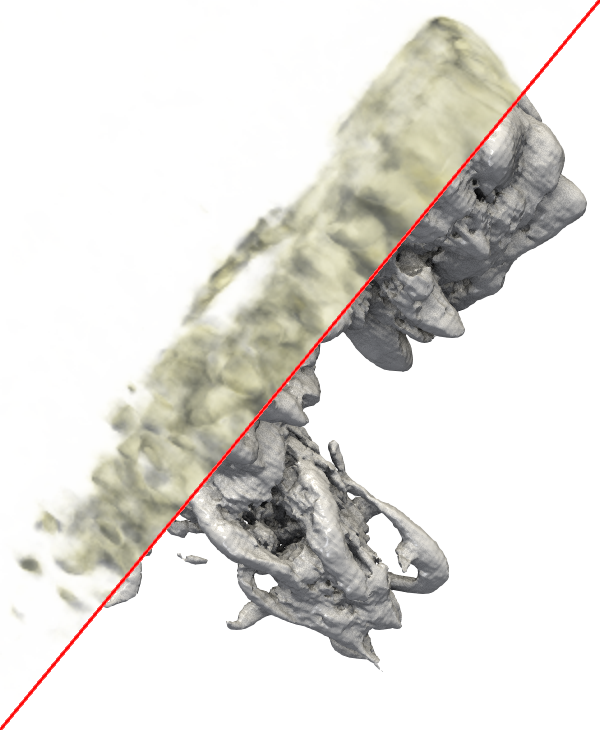}&
  \includegraphics[width=0.1875\linewidth]{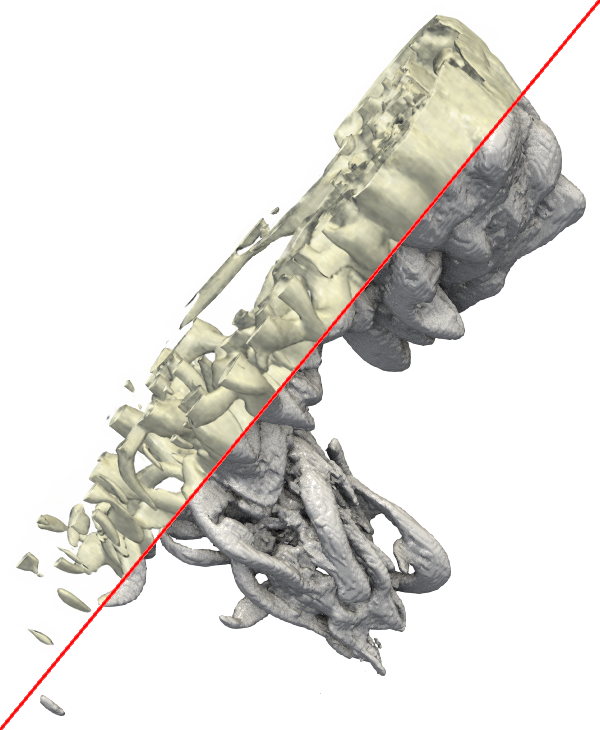}&
  \includegraphics[width=0.1875\linewidth]{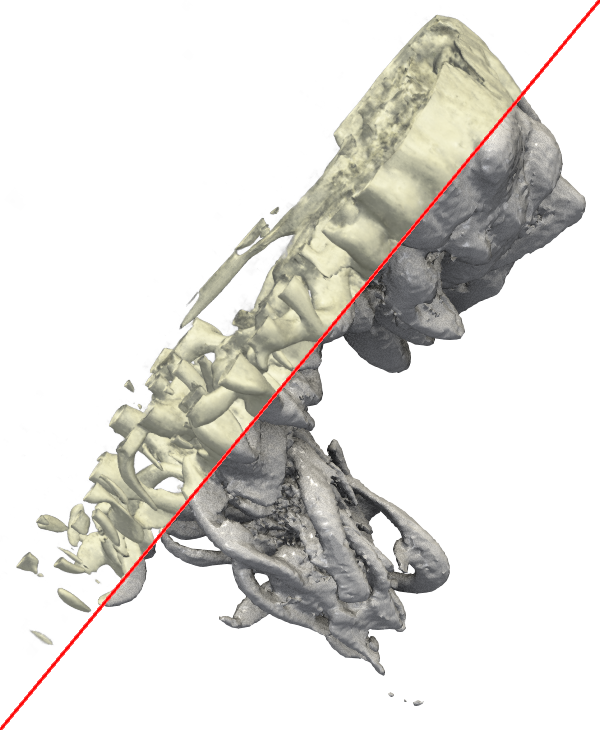}&
  \includegraphics[width=0.1875\linewidth]{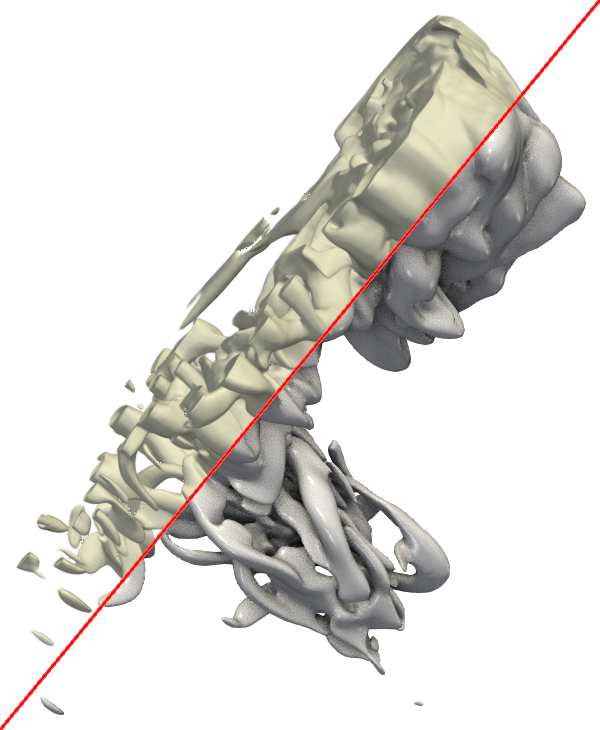}&
  \includegraphics[width=0.1875\linewidth]{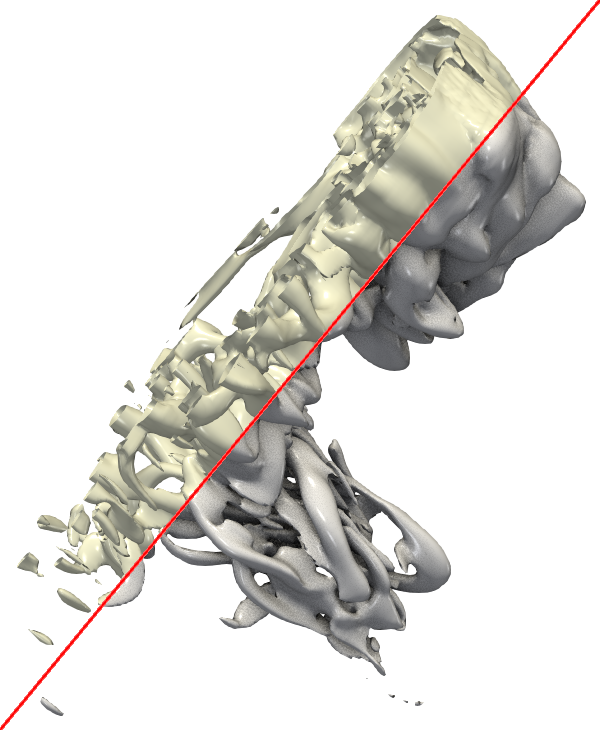}\\
  \includegraphics[width=0.1875\linewidth]{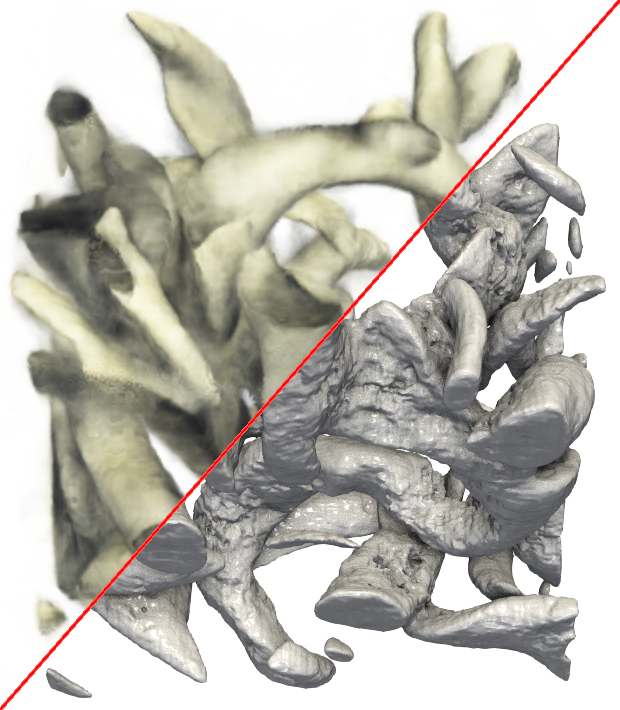}&
  \includegraphics[width=0.1875\linewidth]{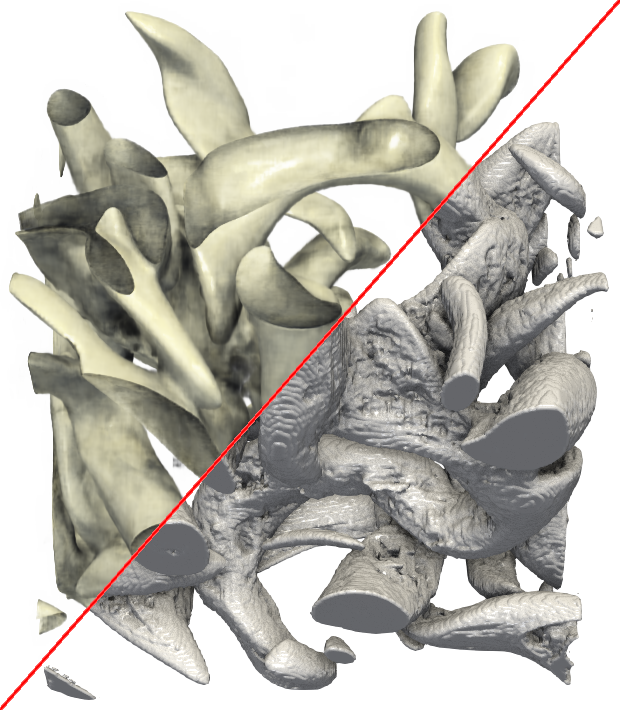}&
  \includegraphics[width=0.1875\linewidth]{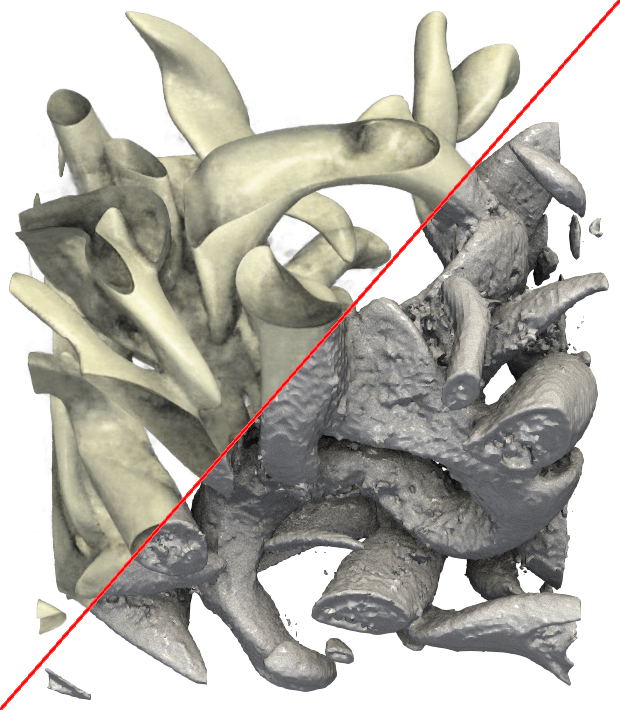}&
  \includegraphics[width=0.1875\linewidth]{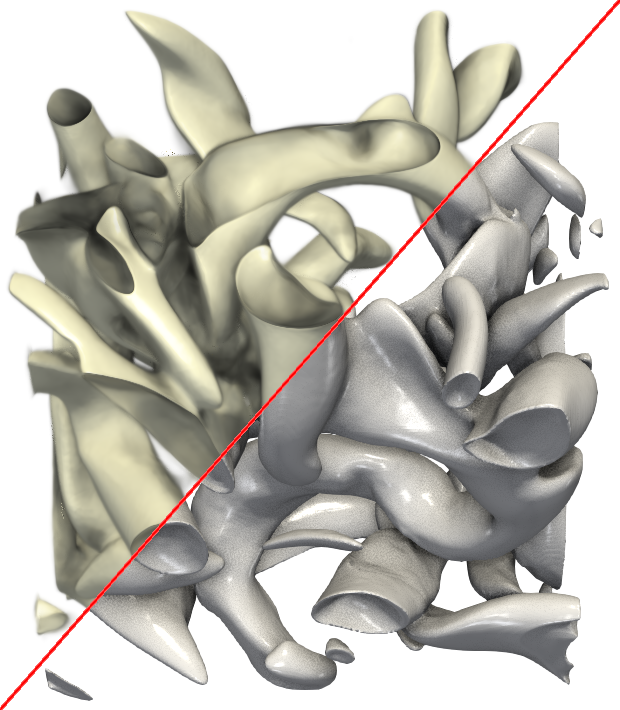}&
  \includegraphics[width=0.1875\linewidth]{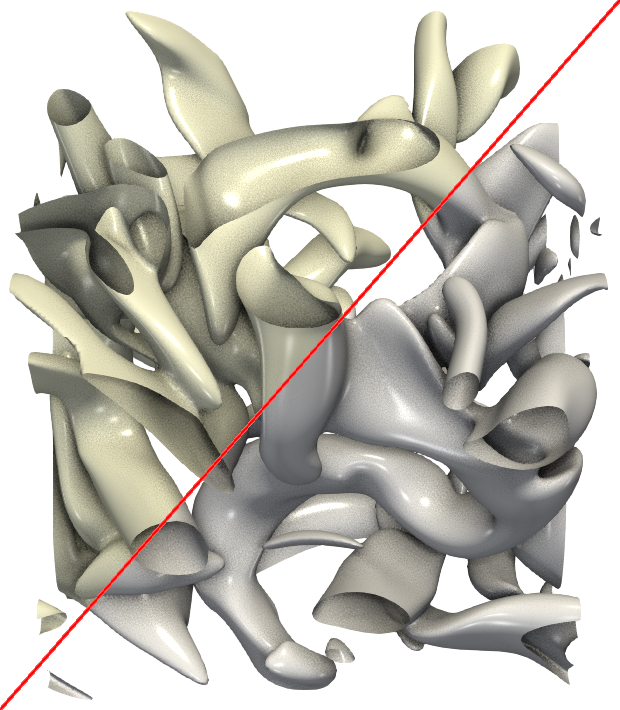}\\
 \mbox{\footnotesize (a)} & \mbox{\footnotesize (b)} & \mbox{\footnotesize (c)} & \mbox{\footnotesize (d)} & \mbox{\footnotesize (e)}
 \end{array}$
 \end{center}
 \vspace{-.25in} 
 \caption{Inferred neural rendering images (upper-left) and rendering images of reconstructed surfaces (lower-right) of Nyx, Tangaroa, and vortex generated by (a) NeRF, (b) TensoRF, (c) Instant-NGP, and (d) NeuS. (e) shows the GT results.}
 \label{fig:comp-nerf}
\end{figure}

\begin{table}[htb]
  \vspace{-0.05in}
  \caption{Comparison of NeuS vs.\ NeRF methods. Average PSNR (dB) and LPIPS across all synthesized views, CD between the reconstructed and GT surfaces, training time (TT, in hours), and model size (MS, in MB). The best ones are highlighted in bold.}
  \vspace{-0.05in}
  \centering
  \resizebox{\columnwidth}{!}{
  \begin{tabular}{c|c|ccc|cc}
  dataset & method & PSNR$\uparrow$ & LPIPS$\downarrow$ & CD$\downarrow$ & TT$\downarrow$ & MS$\downarrow$ \\ \hline
  \multirow{4}{*}{Nyx} & NeRF & 24.20 & 0.224 & $11.32+\mathbf{0.49}$ & 11.67 & 13.69 \\
   & TensoRF & 26.89 & 0.090 & $\mathbf{4.19}+8.14$ & 0.68 & 66.90 \\
   & Instant-NGP & \bf{27.82} & \bf{0.075} & $7.09+788.12$ & \bf{0.05} & 23.80 \\
   & NeuS & 23.74 & 0.157 & $5.43+0.58$ & 11.01 & \bf{11.55} \\ \hline
  \multirow{4}{*}{Tangaroa} & NeRF & 25.83 & 0.145 & $3.38+1.31$ & 11.31 & 13.69 \\
   & TensoRF & \bf{32.08} & \bf{0.029} & $\mathbf{0.75}+1.00$ & 0.69 & 69.16 \\
   & Instant-NGP & 31.73 & 0.034 & $0.78+90.54$ & \bf{0.06} & 23.80 \\
   & NeuS & 30.13 & 0.039 & $4.20+\mathbf{0.55}$ & 10.98 & \bf{11.55} \\ \hline
  \multirow{4}{*}{vortex} & NeRF & 23.21 & 0.220 & $1.40+2.37$ & 11.72 & 13.69 \\
   & TensoRF & \bf{26.55} & 0.101 & $\mathbf{0.26}+3.84$ & 0.73 & 67.13 \\
   & Instant-NGP & 26.54 & \bf{0.096} & $0.30+427.88$ & \bf{0.06} & 24.00 \\
   & NeuS & 24.65 & 0.146 & $0.29+\mathbf{1.14}$ & 10.30 & \bf{11.55} \\
  \end{tabular}
  }
  \label{tab:comp-nerf}
 \vspace{-0.1in}
 \end{table}

\vspace{-0.05in}
\subsection{Comparison Results}

{\bf NeuS vs.\ NeRF methods.}
Figure~\ref{fig:comp-nerf} shows the qualitative results. For space-saving, we display the inferred neural rendering images on the upper-left side and the rendering images of reconstructed surfaces on the lower-right side. 
For neural rendering, the results indicate that except for the vanilla NeRF, which leads to blurred results, all other methods can generate inferred neural rendering images with reasonably good quality. 
NeRF methods are capable of generating reconstructed surfaces of reasonable quality.
Nonetheless, reconstructing the underlying surfaces is more challenging than inferring novel views. 
We can see that NeRF-generated surfaces exhibit noticeable noise, especially for datasets with larger areas of even surfaces like Tangaroa and vortex.
In contrast, NeuS delivers surfaces with a significantly higher degree of smoothness, owing to the integration of Eikonal regularization. 
This technique effectively aligns the normals across the surface, enhancing the smoothness and coherence of the reconstructed surface.

The quantitative results presented in Table~\ref{tab:comp-nerf} reveal that NeRF and NeuS do not yield neural renderings as detailed as the other two methods due to their fully implicit design. 
Nonetheless, NeuS, benefiting from its SDF representation, consistently maintains a low value of the second CD term across all datasets. 
This indicates that while some parts of the original surface may be missing, the reconstructed surface does not generate excessive extraneous elements. 
Conversely, for Instant-NGP, MHE is responsible for introducing outliers significantly apart from the actual surface, resulting in an exceptionally high second CD term.

Regarding the training speed, the hybrid architectures of TensoRF and Instant-NGP demonstrate a considerable acceleration over the fully implicit models of NeRF and NeuS. 
This advantage, however, incurs an increased model size for storing explicit representations.
Between Instant-NGP and TensoRF, Instant-NGP is more than 10$\times$ faster to train than TensoRF. 

In summary, this group analysis suggests that while occupancy-based surface representations employed by NeRF are somewhat effective, the SDF-based representation showcases superior efficacy in reconstructing smooth and consistent surfaces.

\begin{figure}[htb]
 \vspace{-.1in} 
  \begin{center}
  $\begin{array}{c@{\hspace{0.02in}}c@{\hspace{0.02in}}c@{\hspace{0.02in}}c@{\hspace{0.02in}}c}
  \includegraphics[width=0.1875\linewidth]{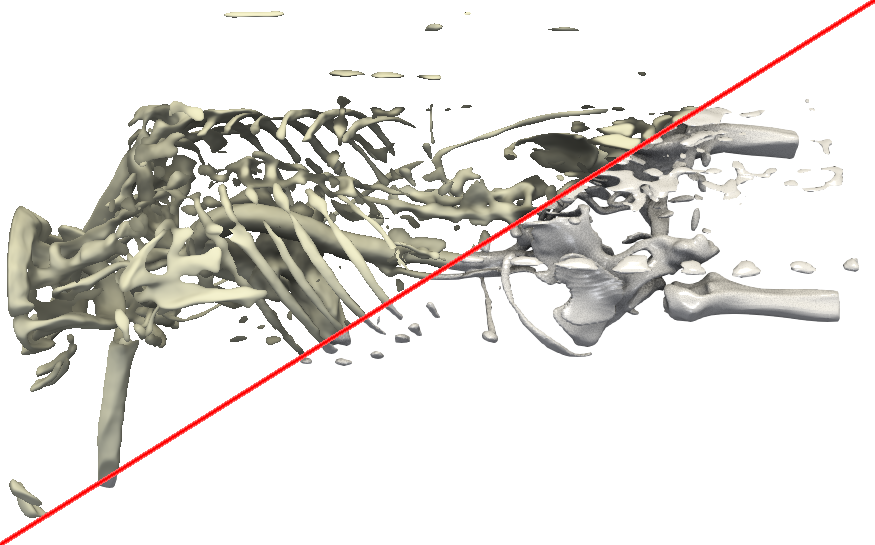}&
  \includegraphics[width=0.1875\linewidth]{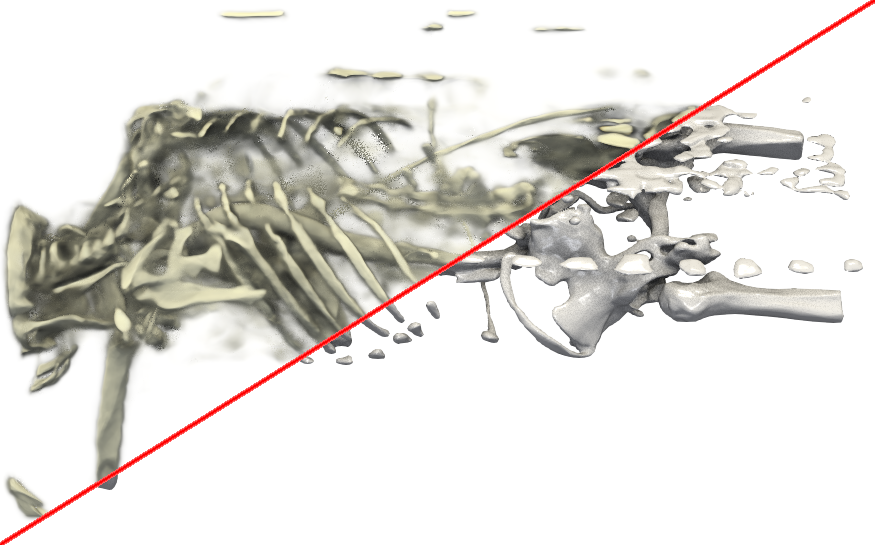}&
  \includegraphics[width=0.1875\linewidth]{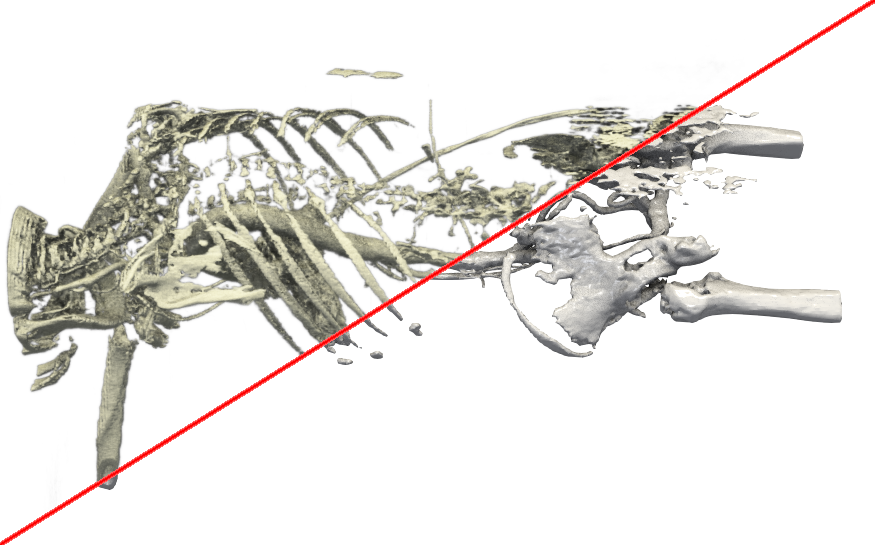}&
  \includegraphics[width=0.1875\linewidth]{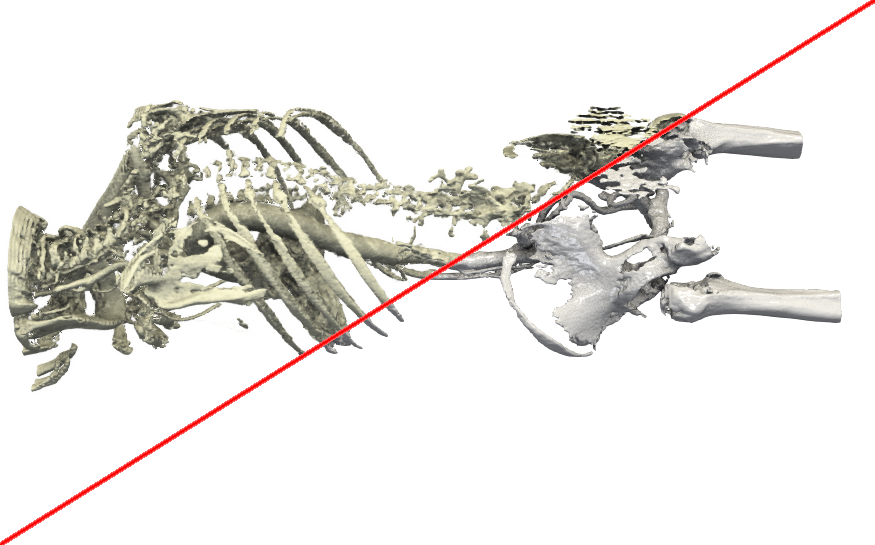}&
  \includegraphics[width=0.1875\linewidth]{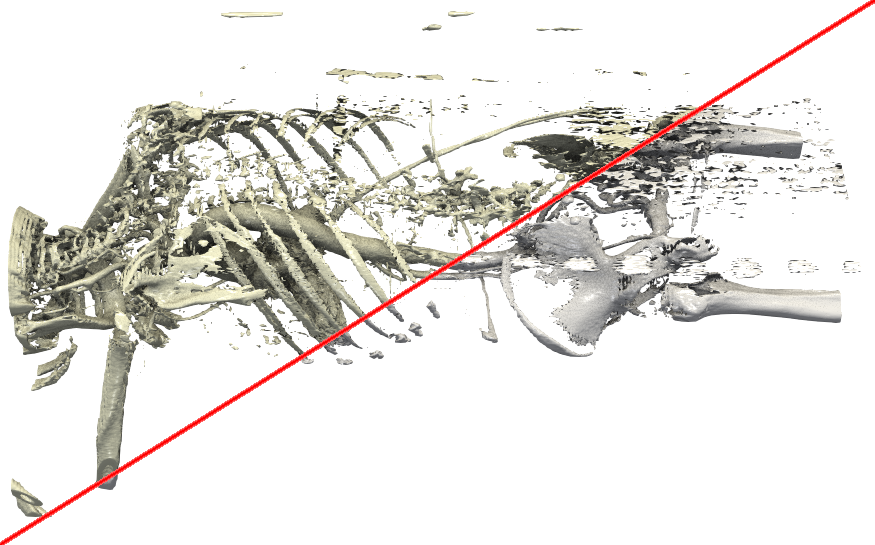}\\
  \includegraphics[width=0.1875\linewidth]{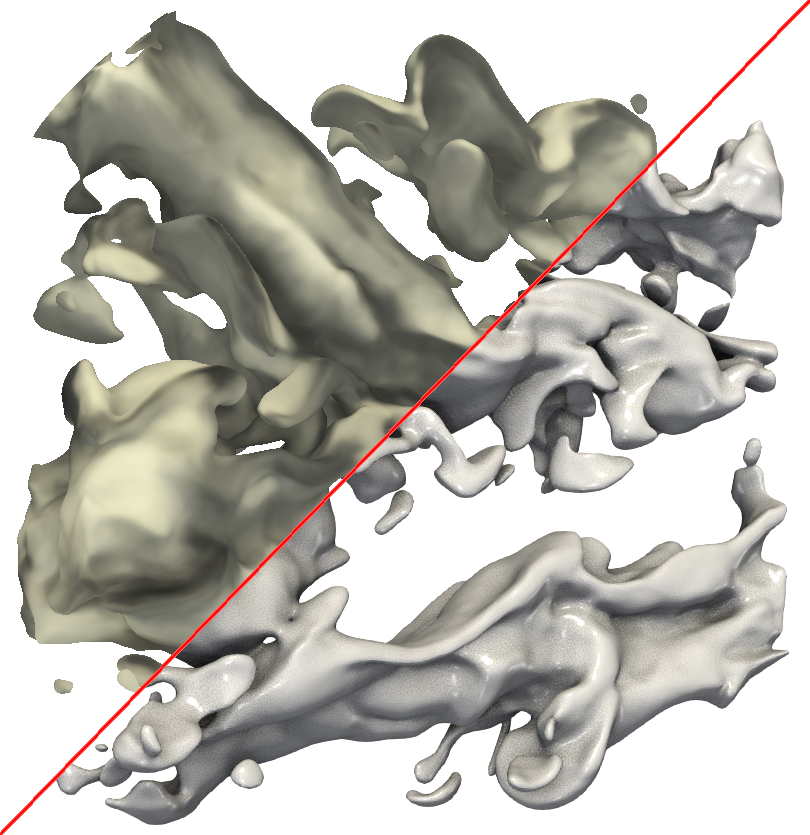}&
  \includegraphics[width=0.1875\linewidth]{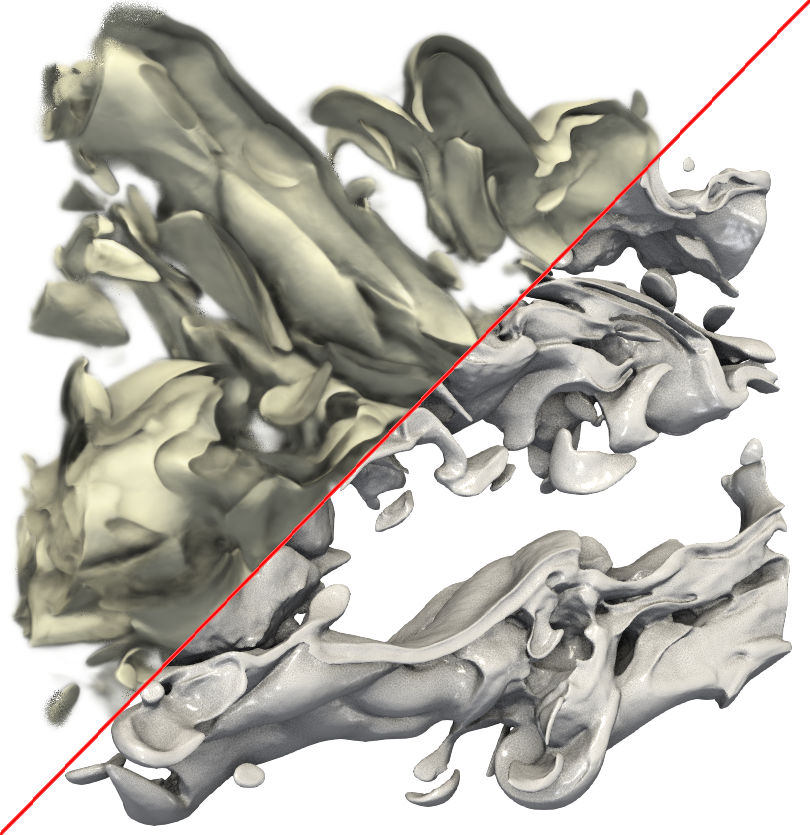}&
  \includegraphics[width=0.1875\linewidth]{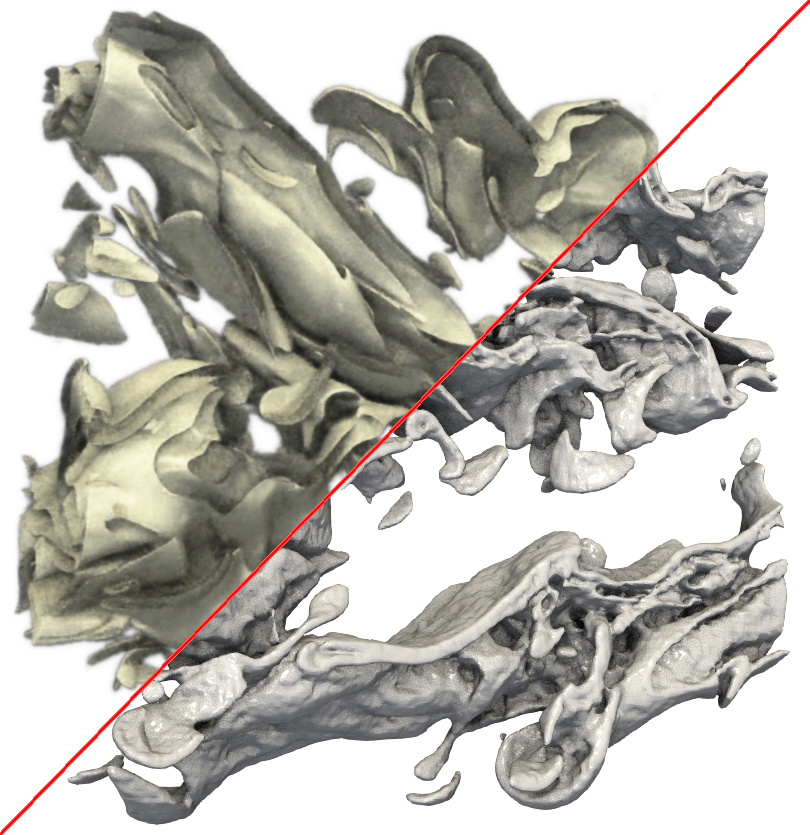}&
  \includegraphics[width=0.1875\linewidth]{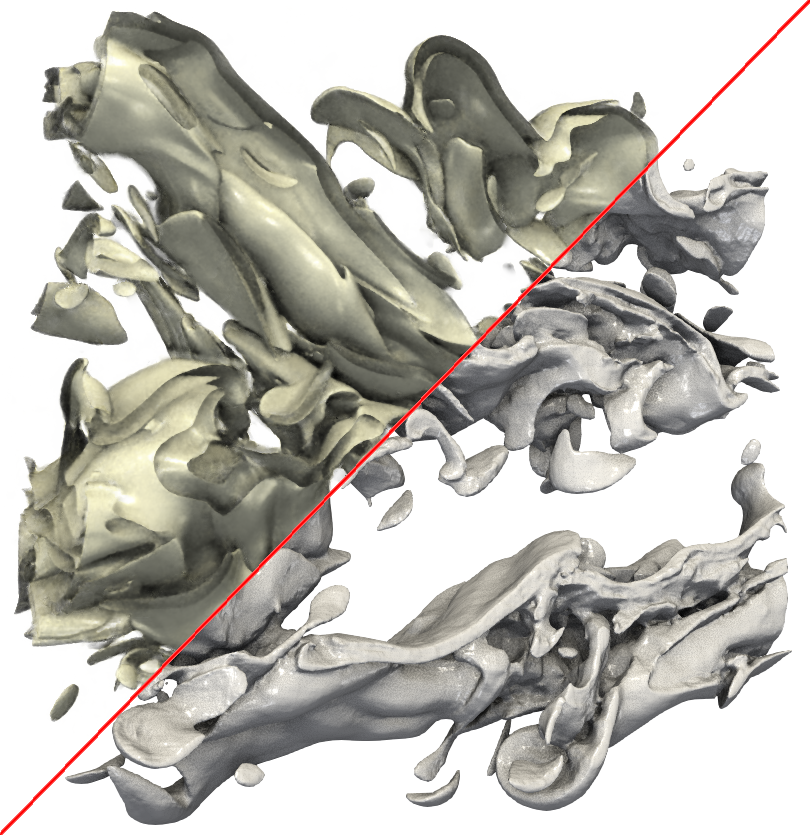}&
  \includegraphics[width=0.1875\linewidth]{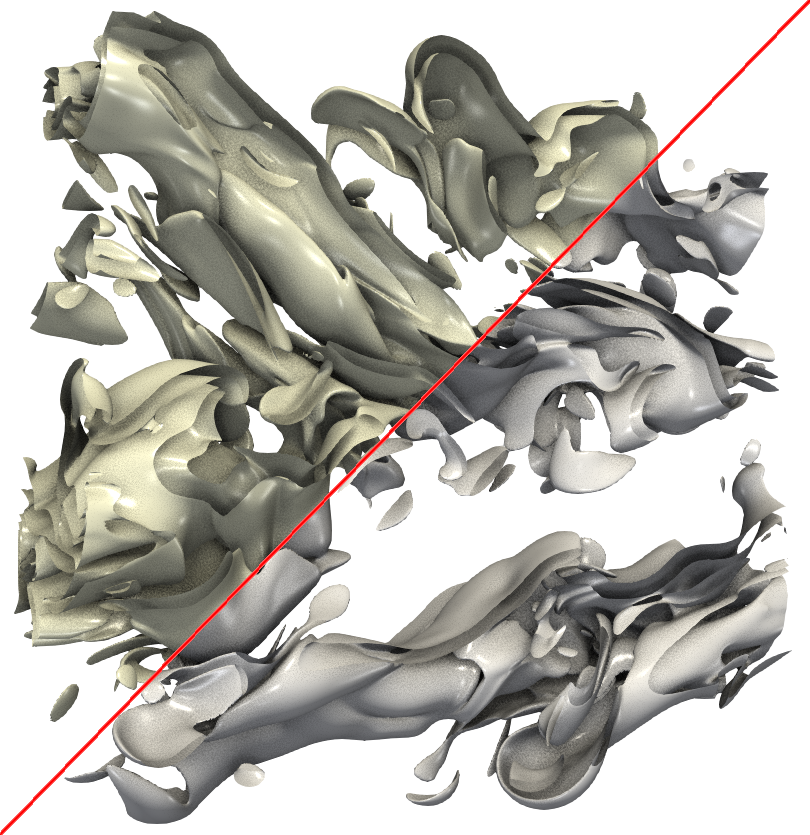}\\
  \includegraphics[width=0.1875\linewidth]{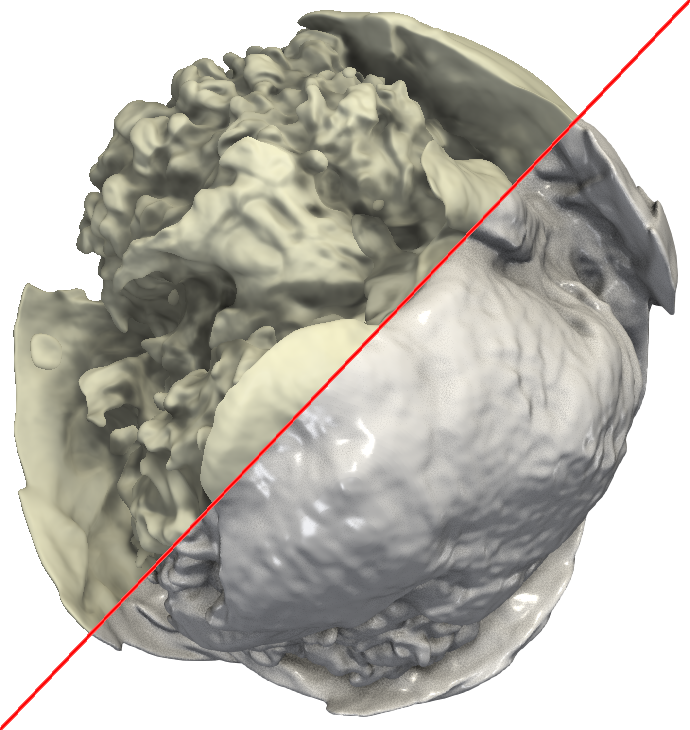}&
  \includegraphics[width=0.1875\linewidth]{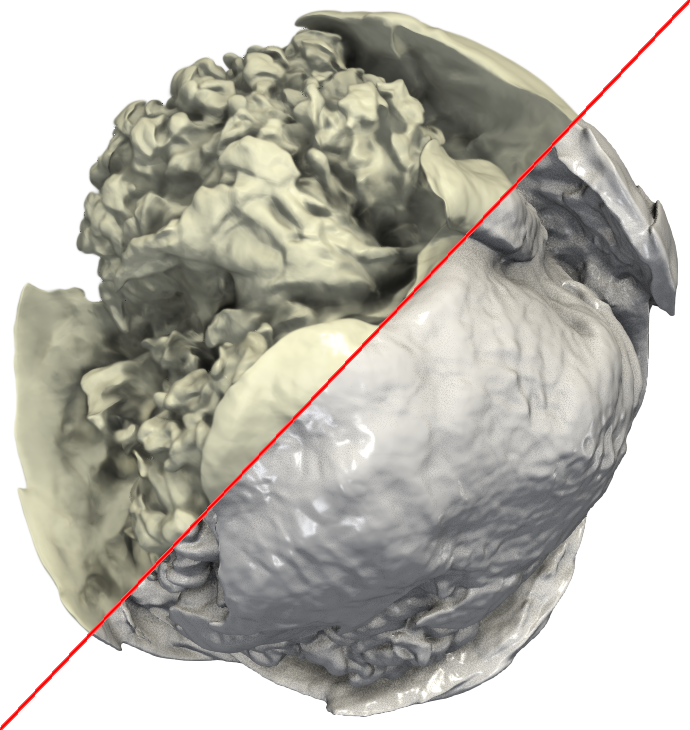}&
  \includegraphics[width=0.1875\linewidth]{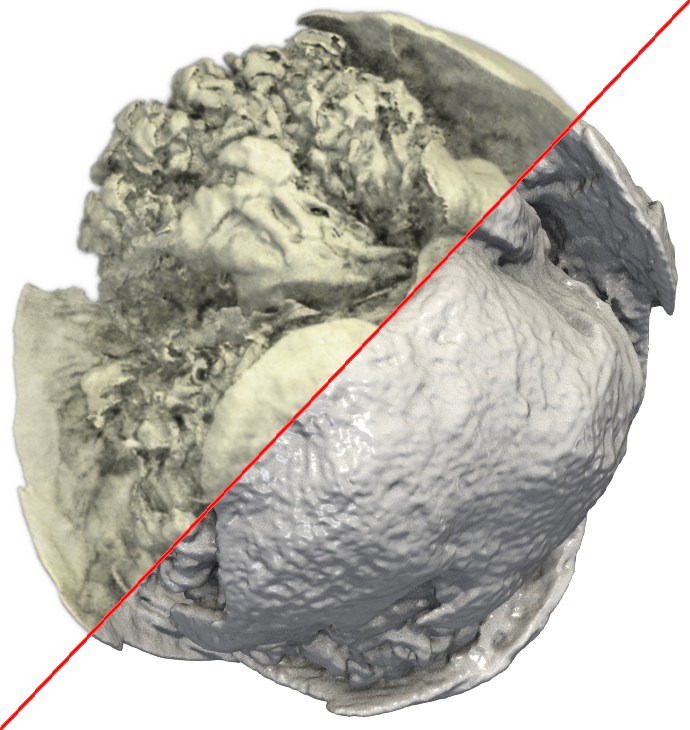}&
  \includegraphics[width=0.1875\linewidth]{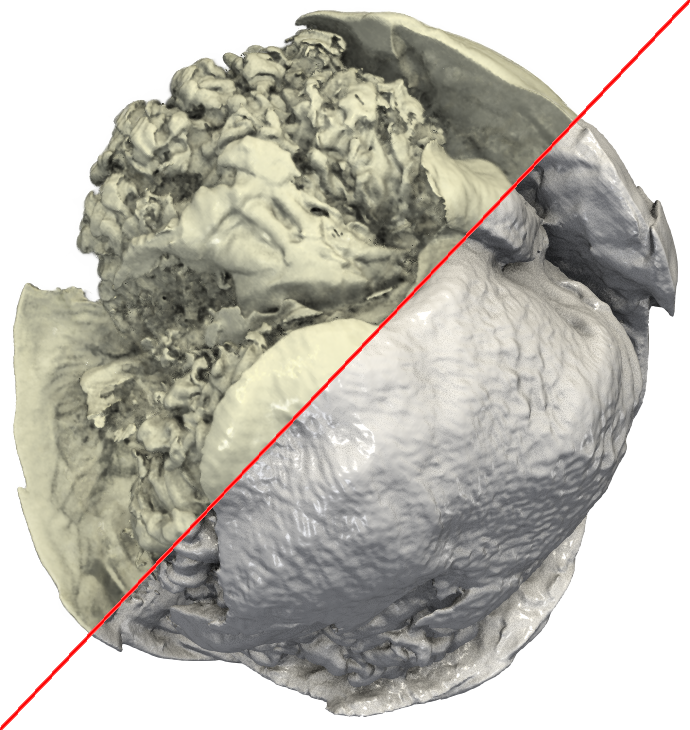}&
  \includegraphics[width=0.1875\linewidth]{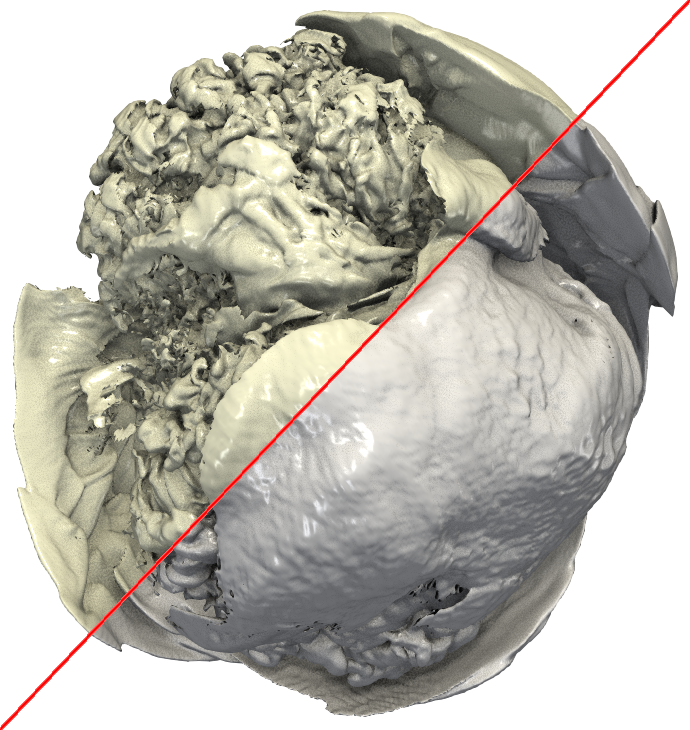}\\
 \mbox{\footnotesize (a)} & \mbox{\footnotesize (b)} & \mbox{\footnotesize (c)} & \mbox{\footnotesize (d)} & \mbox{\footnotesize (e)}
 \end{array}$
 \end{center}
 \vspace{-.25in} 
 \caption{Inferred neural rendering images (upper-left) and rendering images of reconstructed surfaces (lower-right) of aorta, combustion, and supernova generated by (a) IDR, (b) NeuS, (c) NeuS2, and (d) Neuralangelo. (e) shows the GT results.}
 \label{fig:comp-closed}
\end{figure}

\begin{figure}[htb]
 \vspace{-.1in} 
  \begin{center}
  $\begin{array}{c@{\hspace{0.02in}}c@{\hspace{0.02in}}c@{\hspace{0.02in}}c@{\hspace{0.02in}}c}
  \includegraphics[width=0.1875\linewidth]{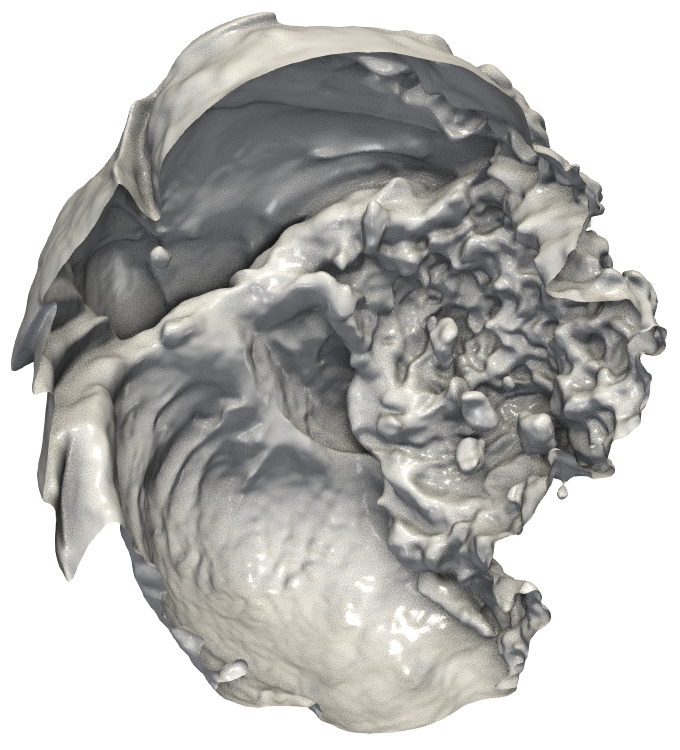}&
  \includegraphics[width=0.1875\linewidth]{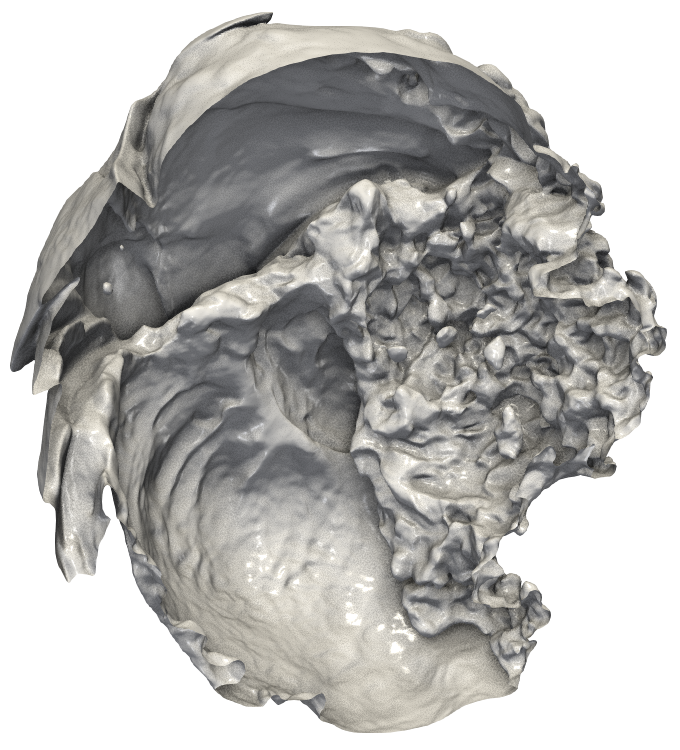}&
  \includegraphics[width=0.1875\linewidth]{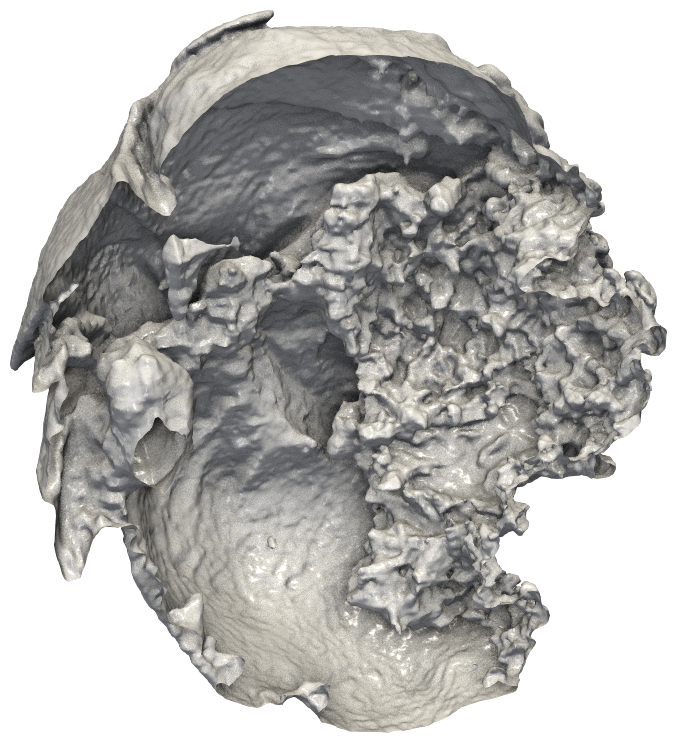}&
  \includegraphics[width=0.1875\linewidth]{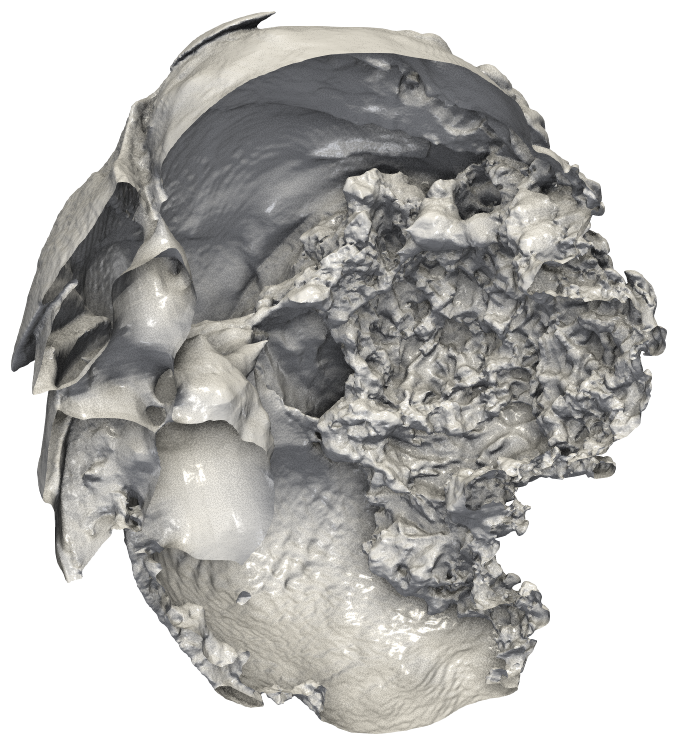}&
  \includegraphics[width=0.1875\linewidth]{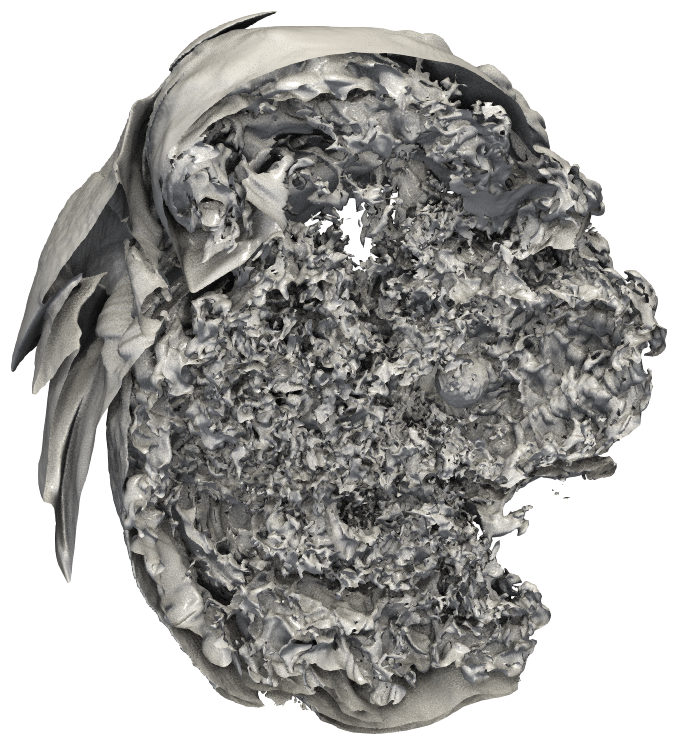}\\
 \mbox{\footnotesize (a)} & \mbox{\footnotesize (b)} & \mbox{\footnotesize (c)} & \mbox{\footnotesize (d)} & \mbox{\footnotesize (e)}
 \end{array}$
 \end{center}
 \vspace{-.25in} 
 \caption{Cut-through rendering of reconstructed supernova surfaces generated by (a) IDR, (b) NeuS, (c) NeuS2, and (d) Neuralangelo. (e) shows the GT results.}
 \label{fig:comp-supernova-clip}
\end{figure}

\begin{table}[htb]
  \vspace{-0.05in}
  \caption{Comparison of neural implicit closed surface methods.}
  \vspace{-0.05in}
  \centering
  \resizebox{\columnwidth}{!}{
  \begin{tabular}{c|c|ccc|cc}
  dataset & method & PSNR$\uparrow$ & LPIPS$\downarrow$ & CD$\downarrow$ & TT$\downarrow$ & MS$\downarrow$ \\ \hline
  \multirow{4}{*}{aorta} & IDR & 20.06 & \bf{0.100} & $\mathbf{24.70}+19.80$ & 4.75 & \bf{11.13} \\
   & NeuS & 20.48 & 0.138 & $35.25+12.18$ & 10.89 & 11.55 \\
   & NeuS2 & \bf{23.22} & \bf{0.100} & $127.03+\mathbf{1.76}$ & \bf{0.13} & 24.15 \\
   & Neuralangelo & 22.27 & 0.117 & $433.59+2.07$ & 23.87 & 1396.81 \\ \hline
  \multirow{4}{*}{combustion} & IDR & 20.19 & 0.210 & $74.61+29.20$ & 5.03 & \bf{11.13} \\
   & NeuS & 22.70 & 0.192 & $32.34+18.37$ & 11.01 & 11.55 \\
   & NeuS2 & 24.56 & 0.161 & $\mathbf{21.52}+12.37$ & \bf{0.12} & 24.15 \\
   & Neuralangelo & \bf{25.05} & \bf{0.114} & $24.21+\mathbf{12.35}$ & 23.81 & 1396.81 \\ \hline
  \multirow{4}{*}{supernova} & IDR & 25.12 & 0.198 & $692.26+9.54$ & 5.61 & \bf{11.13} \\
   & NeuS & 26.08 & 0.179 & $729.48+\mathbf{9.17}$ & 10.58 & 11.55 \\
   & NeuS2 & 26.63 & 0.162 & $334.14+9.44$ & \bf{0.13} & 24.15 \\
   & Neuralangelo & \bf{27.84} & \bf{0.112} & $\mathbf{281.26}+10.87$ & 23.73 & 1396.81 \\ 
  \end{tabular}
  }
  \label{tab:comp-closed}
 \vspace{-0.1in}
 \end{table}

{\bf Neural implicit closed surface methods.} 
Analyzing the results in Figure~\ref{fig:comp-closed} reveals that IDR and NeuS tend to produce blurry images, in contrast to NeuS2 and Neuralangelo, which, utilizing MHE, can generate more detailed neural renderings.
Compared to NeuS2, the neural renderings produced by Neuralangelo are characterized by improved smoothness and enhanced specular highlights.
However, within the aorta dataset, NeuS2 and Neuralangelo exhibit missing components in the reconstructed surfaces, stemming from omitted isolated parts in voxels at the lower-resolution hash grids.

Table~\ref{tab:comp-closed} indicates that Neuralangelo achieves superior overall neural rendering quality. 
However, it is outperformed by NeuS2 for the aorta dataset due to the absence of the prominent collarbone and other details.
IDR records the lowest overall CD for the aorta dataset, but its high second CD term implies the generation of inaccurate pieces.
NeuS2 and Neuralangelo significantly surpass the other methods in neural rendering and surface reconstruction of the combustion and supernova datasets, which predominantly consist of dense surfaces. 
Upon closer examination, we find that the supernova surface exhibits a complex inner structure yet is heavily occluded from most camera views. 
To illustrate this, we slice the surface in half to expose its interior, as shown in Figure~\ref{fig:comp-supernova-clip}. 
The results indicate that all methods fail to reconstruct the intricate hidden details, leading to a notably high first CD term as shown in Table~\ref{tab:comp-closed}.
Among these methods, Neuralangelo handles this scenario slightly better than its counterparts.

Although IDR and NeuS share similar model architectures, IDR trains faster due to its use of surface instead of volume rendering.
NeuS2, utilizing tiny MLPs optimized in the CUDA framework, achieves a speed increase of approximately 100$\times$ compared to NeuS. 
Neuralangelo, with enormous hash grids and more training iterations, takes twice as long as NeuS.

The model sizes of both NeuS2 and Neuralangelo largely depend on their MHE configurations.
Neuralangelo incorporates 16 levels of MHE, with resolution from $2^5$ to $2^{11}$, and each resolution uses $2^{22}$ hash entries.
In contrast, NeuS2 employs 14 levels, with resolution from $2^4$ to $2^{11}$, and allocates $2^{19}$ hash entries per level.

Given the above findings, we recommend using NeuS2 for reconstructing closed surfaces due to its extraordinary efficiency and high-quality reconstruction.

\begin{figure}[htb]
 \vspace{-.1in} 
  \begin{center}
  $\begin{array}{c@{\hspace{0.02in}}c@{\hspace{0.02in}}c@{\hspace{0.02in}}c@{\hspace{0.02in}}c}
  \includegraphics[width=0.1875\linewidth]{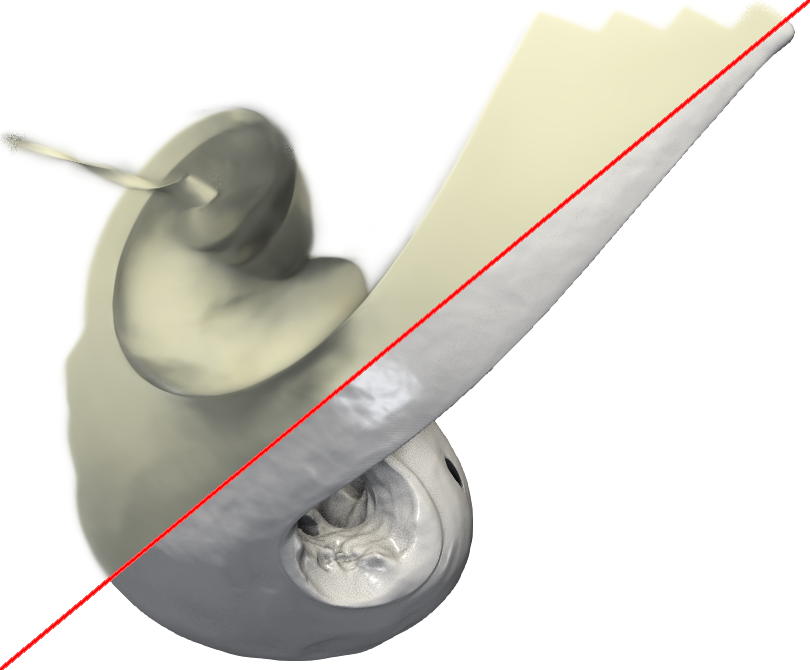}&
  \includegraphics[width=0.1875\linewidth]{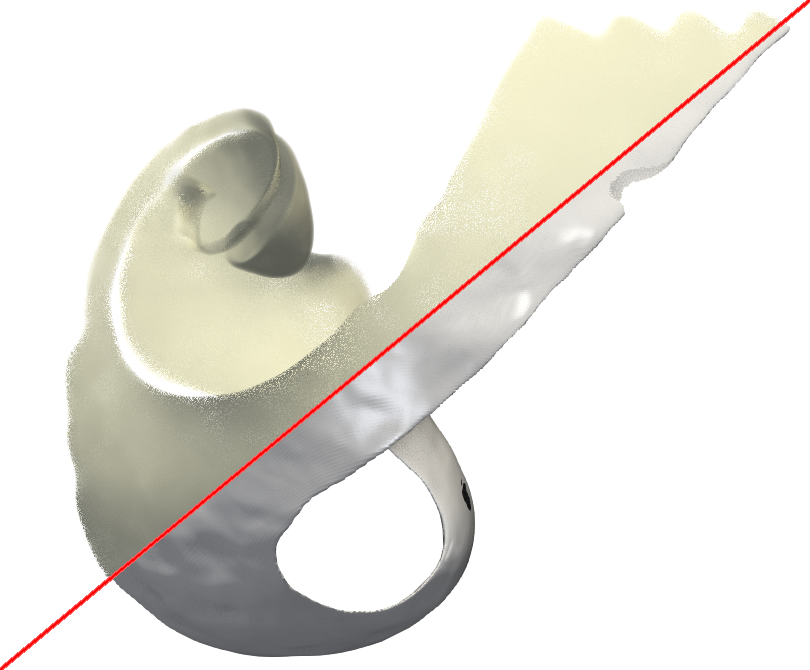}&
  \includegraphics[width=0.1875\linewidth]{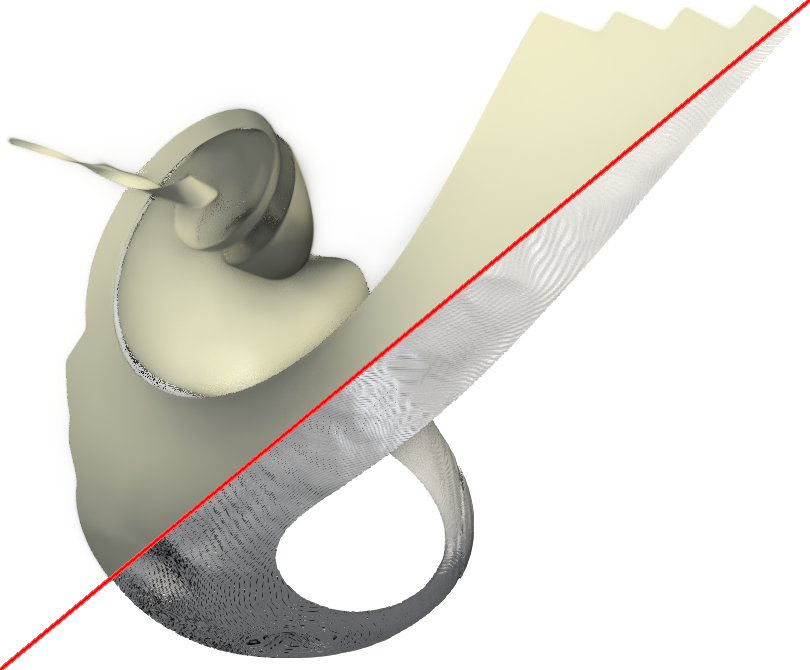}&
  \includegraphics[width=0.1875\linewidth]{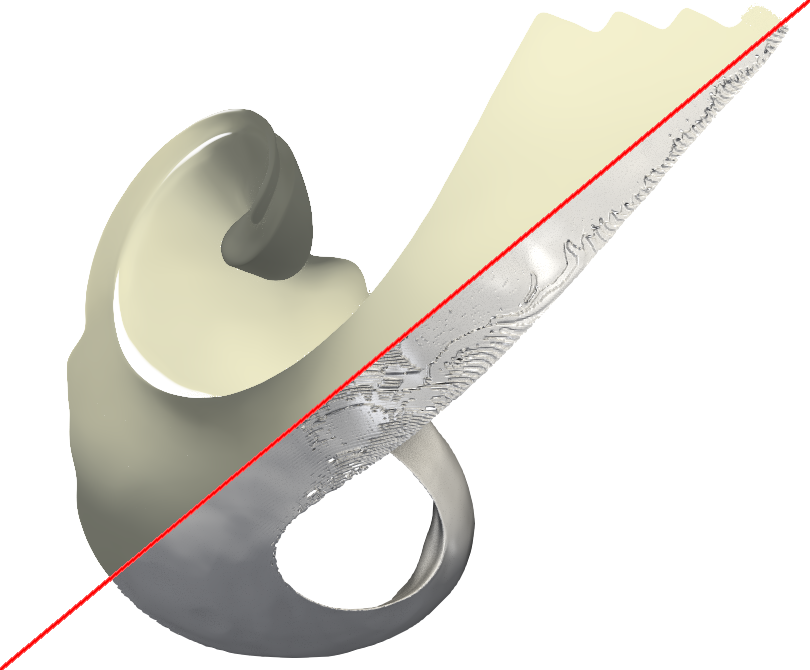}&
  \includegraphics[width=0.1875\linewidth]{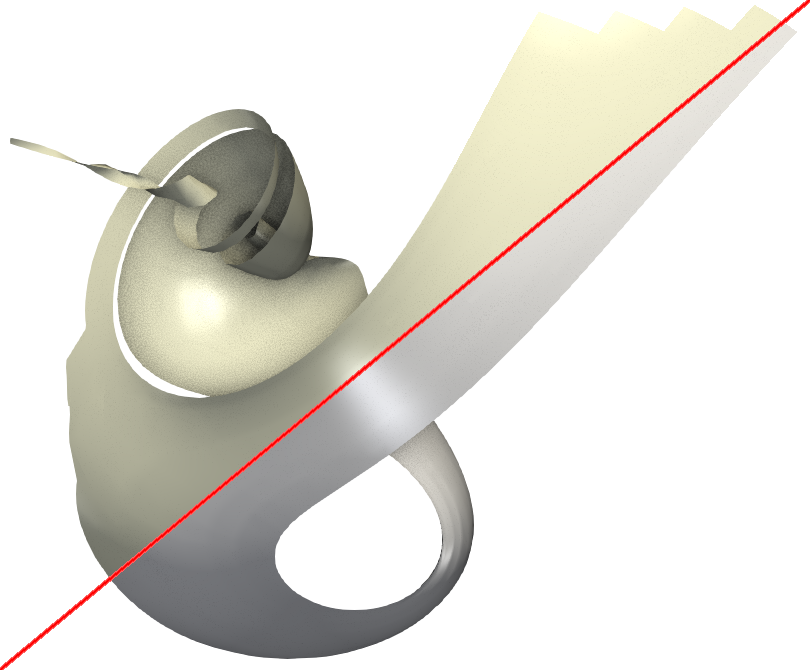}\\
  \includegraphics[width=0.1875\linewidth]{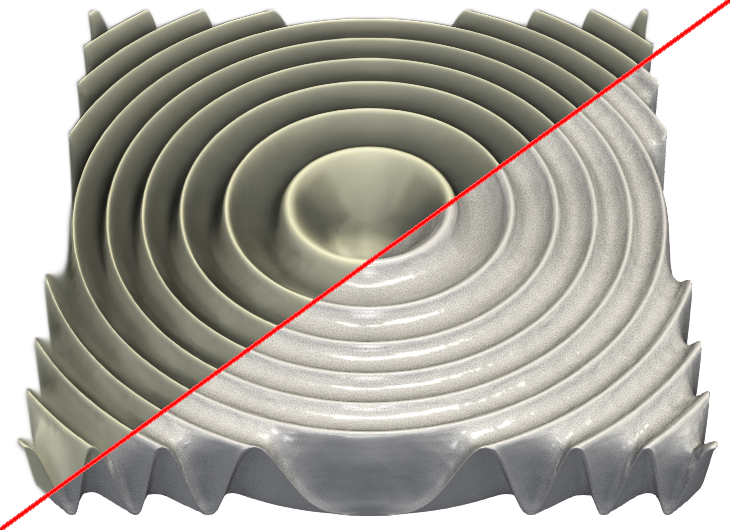}&
  \includegraphics[width=0.1875\linewidth]{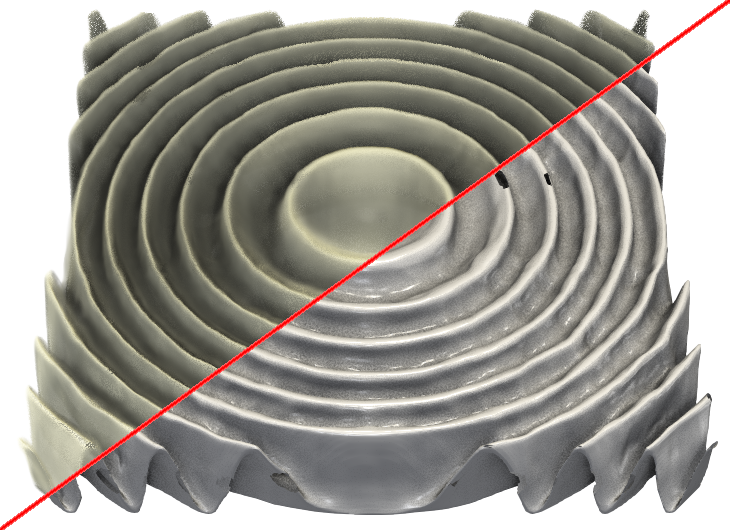}&
  \includegraphics[width=0.1875\linewidth]{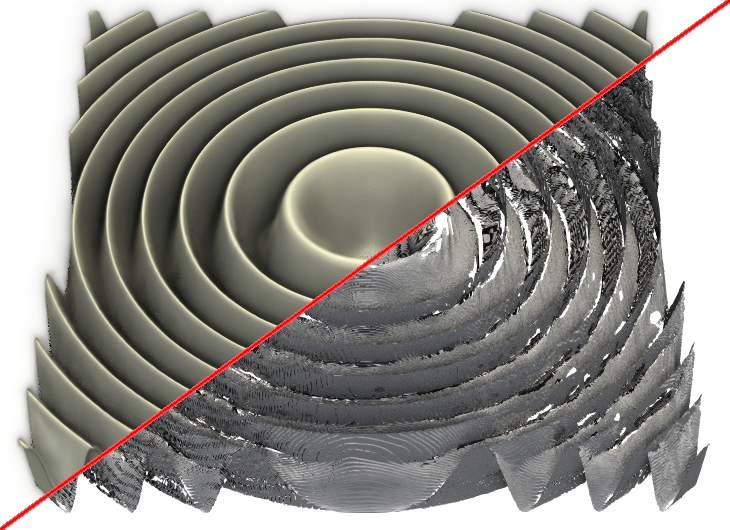}&
  \includegraphics[width=0.1875\linewidth]{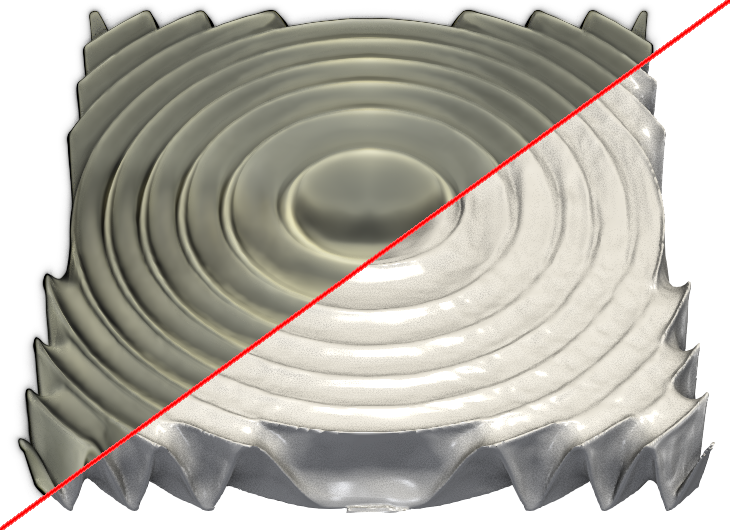}&
  \includegraphics[width=0.1875\linewidth]{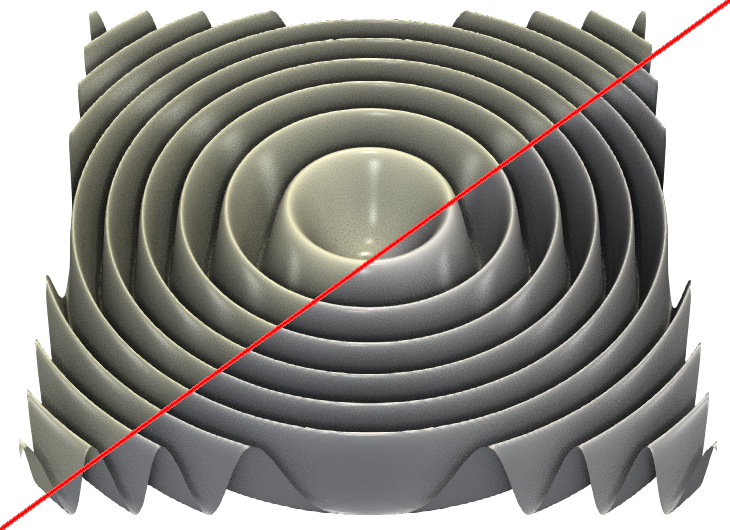}\\
  \includegraphics[width=0.1875\linewidth]{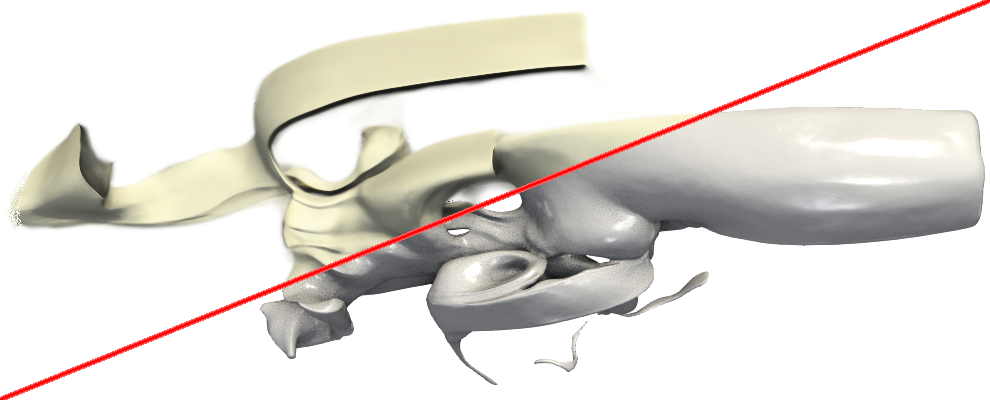}&
  \includegraphics[width=0.1875\linewidth]{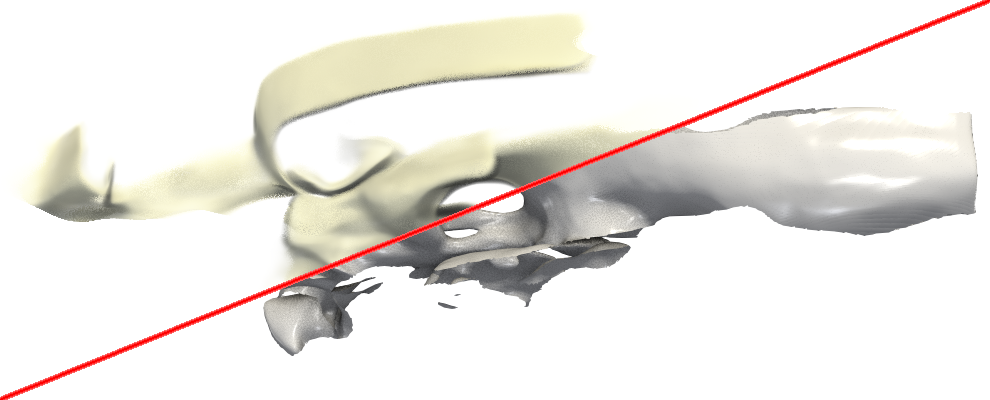}&
  \includegraphics[width=0.1875\linewidth]{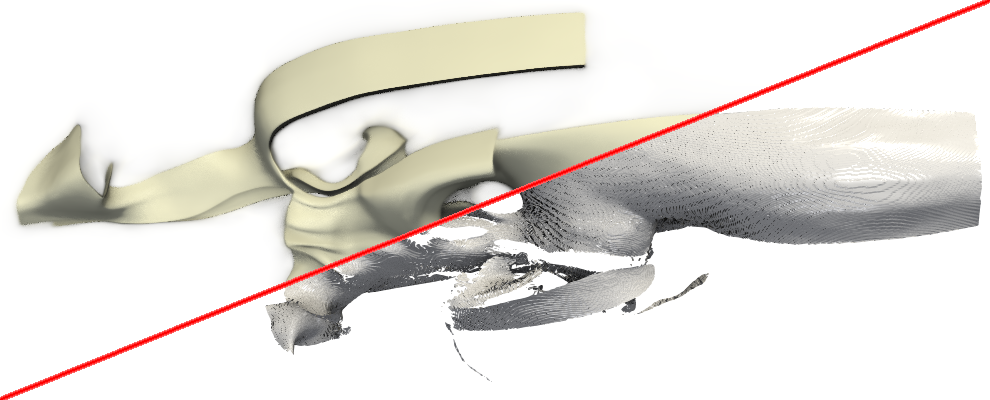}&
  \includegraphics[width=0.1875\linewidth]{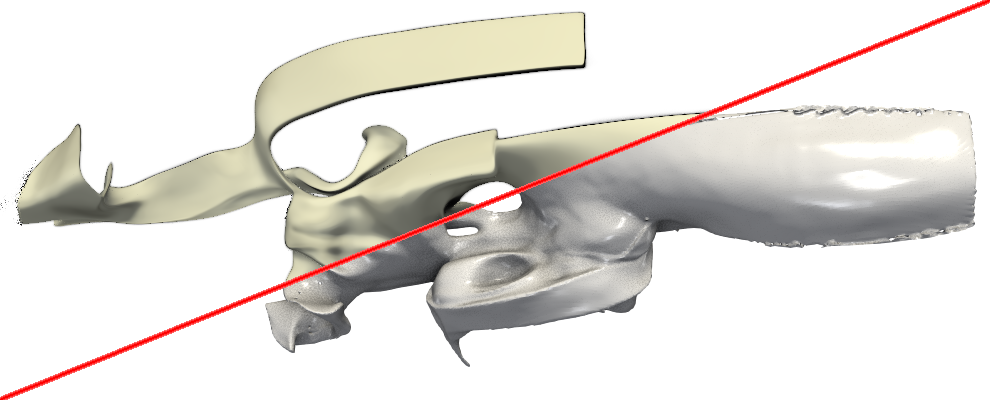}&
  \includegraphics[width=0.1875\linewidth]{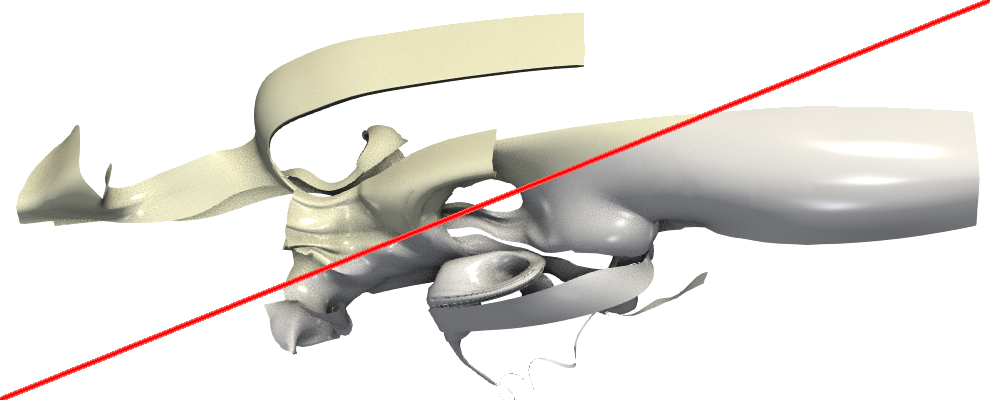}\\
 \mbox{\footnotesize (a)} & \mbox{\footnotesize (b)} & \mbox{\footnotesize (c)} & \mbox{\footnotesize (d)} & \mbox{\footnotesize (e)}
 \end{array}$
 \end{center}
 \vspace{-.25in} 
 \caption{Inferred neural rendering images (upper-left) and rendering images of reconstructed surfaces (lower-right) of five critical points, Marschner-Lobb, and solar plume generated by (a) NeuS, (b) NeAT, (c) NeUDF, and (d) NeuralUDF. (e) shows the GT results.}
 \label{fig:comp-open}
\end{figure}

\begin{table}[htb]
  \vspace{-0.05in}
  \caption{Comparison of neural implicit open surface methods.}
  \vspace{-0.05in}
  \centering
  \resizebox{\columnwidth}{!}{
  \begin{tabular}{c|c|ccc|cc}
  dataset & method & PSNR$\uparrow$ & LPIPS$\downarrow$ & CD$\downarrow$ & TT$\downarrow$ & MS$\downarrow$ \\ \hline
  \multirow{4}{*}{five critical points} & NeuS & 23.60 & 0.087 & $0.11+9.52$ & \bf{11.12} & 11.55 \\
   & NeAT & \bf{26.99} & 0.102 & $0.64+0.41$ & 33.10 & \bf{9.50} \\
   & NeUDF & 26.89 & \bf{0.040} & $\mathbf{0.05}+\mathbf{0.31}$ & 16.92 & 11.55 \\
   & NeuralUDF & 26.30 & 0.049 & $1.57+0.57$ & 20.64 & 10.23 \\ \hline
  \multirow{4}{*}{Marschner-Lobb} & NeuS & \bf{25.51} & 0.084 & $\mathbf{0.35}+1.03$ & \bf{10.77} & 11.55 \\
   & NeAT & 24.36 & \bf{0.074} & $0.40+0.91$ & 33.99 & \bf{9.50} \\
   & NeUDF & 25.41 & 0.084 & $1.33+\mathbf{0.87}$ & 17.41 & 11.55 \\
   & NeuralUDF & 21.69 & 0.104 & $0.40+1.08$ & 20.04 & 10.23 \\ \hline
  \multirow{4}{*}{solar plume} & NeuS & 26.04 & 0.079 & $0.26+1.24$ & \bf{10.68} & 11.55 \\
   & NeAT & 22.50 & 0.113 & $\mathbf{0.16}+0.36$ & 33.22 & \bf{9.50} \\
   & NeUDF & \bf{26.92} & \bf{0.063} & $0.18+\mathbf{0.21}$ & 16.70 & 11.55 \\
   & NeuralUDF & 25.39 & 0.079 & $0.25+0.58$ & 19.36 & 10.23 \\ 
  \end{tabular}
  }
  \label{tab:comp-open}
 \vspace{-0.1in}  
\end{table}

{\bf NeuS vs.\ neural implicit open surface methods.}
Table~\ref{tab:comp-open} shows that methods based on SDF and UDF exhibit comparable neural rendering performance.
However, NeAT and NeuralUDF demonstrate less consistency in their results than NeuS and NeUDF.
According to Figure~\ref{fig:comp-open}, NeAT and NeuralUDF miss the noticeable spiral's tip in their neural renderings for the five critical points dataset.

For stream surfaces of the five critical points and solar plume datasets, NeUDF achieves the lowest overall CD, indicating its most accurately reconstructed surfaces.
The rendering results further validate its ability to reconstruct stream surfaces with high fidelity.

For the isosurface of the Marschner-Lobb dataset, NeUDF stands out as the only method capable of accurately reconstructing the isosurface as a single open surface despite numerous artifacts. 
Contrarily, other methods often yield a watertight surface reconstruction. 
While NeUDF effectively learns the correct surface representation using UDF, improvements in the surface extraction technique are necessary to achieve optimal open surface reconstruction.

\vspace{-0.05in}
\subsection{Summary}

Across these ten methods compared, 
We see significant advantages in using SDF to depict surfaces instead of relying on density fields. 
This approach markedly enhances the precision and smoothness of the reconstructed surfaces. 
Additionally, numerous follow-up works have shown progress over the initial implementation of SDF techniques. 
Although Neuralangelo stands out for its substantial improvements in reconstructing complex surfaces, the increment of training time and model size is fairly substantial. 
In contrast, NeuS2 presents a more feasible option, offering efficient, high-quality surface reconstruction. 
Therefore, for the reconstruction of closed surfaces, we advocate for the adoption of NeuS2.

Regarding the reconstruction of open surfaces, the challenge intensifies. 
NeUDF emerges as a promising candidate, although the way it extracts surfaces still falls short of expectations. 
Other techniques for open surfaces often overlook fine details or inadvertently create a closed, watertight surface with unintended thickness, deviating from the goal of reproducing an open surface. 
Furthermore, the lack of acceleration strategies among these methods for open surfaces leaves room for innovation and improvement.

\vspace{-0.05in}
\section{Concluding Remarks}

Our comparative study shows that current surface reconstruction methods are efficient and can produce high-quality results within minutes when applied to isosurfaces. 
This success is attributed mainly to SDF surface representations and MHE. 
Furthermore, we recognize the promise of UDF for representing isosurfaces and stream surfaces and reconstructing them as open surfaces, as evidenced by the results of NeUDF. 
Despite these encouraging findings, surface reconstruction faces considerable challenges, especially in tackling occlusion from limited camera views and reconstructing minute surface details. 
We outline three potential future research and development directions that could benefit the scientific visualization community. 
{\bf (1) Detail enhancement:} 
To improve the fidelity and minimize errors in reconstructed surfaces, we should develop strategies to capture the fine details more effectively without significantly impacting training time or inflating model size like Neuralangelo. 
This could involve utilizing new sampling methods (such as cone-casting in Mip-NeRF~\cite{Barron-ICCV21}) or optimizing existing ones (especially MHE) to balance detail fidelity with computational efficiency better.
{\bf (2) Efficient UDF representation:}
This direction involves optimizing UDF computations or exploring novel architectures, like MHE or CUDA framework implementation, to efficiently leverage UDF for open and closed surface reconstruction.
{\bf (3) Surface extraction enhancement:} 
To obtain a high-quality open surface that is reliable for simulation or modeling, it is necessary to refine surface extraction techniques for more robust reconstruction of open surface results with enhanced flexibility.
This may include developing new algorithms or modifying existing ones, such as marching cubes, to better accommodate the complexities of open surface reconstruction.

Finally, we invite researchers to utilize our benchmark dataset, 
publicly available at \url{https://www.kaggle.com/datasets/syaond/scivis-surface-dataset/}, to evaluate the performance of their methods.
We hope this will promote advances in the field by providing a standardized basis for comparison, enabling rigorous evaluation of refined surface reconstruction techniques.

\vspace{-0.05in}
\acknowledgments{This research was supported in part by the U.S.\ National Science Foundation through grants IIS-1955395, IIS-2101696, OAC-2104158, and IIS-2401144, and the U.S.\ Department of Energy through grant DE-SC0023145. The authors would like to thank the anonymous reviewers for their insightful comments.}

\vspace{-0.05in}
\bibliographystyle{abbrv-doi}

\bibliography{template-abbv}
\end{document}